\documentclass[fleqn,usenatbib]{mnras}

\usepackage[T1]{fontenc}

\DeclareRobustCommand{\VAN}[3]{#2}
\let\VANthebibliography\thebibliography
\def\thebibliography{\DeclareRobustCommand{\VAN}[3]{##3}\VANthebibliography}

\usepackage{graphicx}	% Including figure files
\usepackage{amsmath}	% Advanced maths commands
\usepackage{amssymb}	% Extra maths symbols

\numberwithin{equation}{section}

\usepackage{newtxtext,newtxmath}
\usepackage{units} % easier units handling
\usepackage{tabto} % allow for tabbing
\renewcommand{\i}{\ensuremath{\text{i}}} % imaginary unit
\newcommand{\e}{\ensuremath{\text{e}}} % Euler number
\newcommand{\pde}[2]{\displaystyle\frac{\partial #1}{\partial #2}} % 1st order partial differential quotient
\newcommand{\ppde}[2]{\displaystyle\frac{\partial^2 #1}{\partial #2^2}} % 2nd order partial differential quotient
\newcommand{\comment}[1]{} % allow for commenting out paragraphs
\DeclareMathOperator{\im}{Im} % imaginary part
\DeclareMathOperator{\arctanh}{arctanh} % area hyperbolic tangent

\title[Time-dependent Rayleigh--Taylor instability]{The time-dependent Rayleigh--Taylor instability in interstellar shells and supershells, including the \textit{eROSITA} bubbles}

\author[M.~M.~Schulreich and D.~Breitschwerdt]{
Michael M.~Schulreich\thanks{E-mail: schulreich@astro.physik.tu-berlin.de}
and Dieter Breitschwerdt
\\
Zentrum f\"ur Astronomie und Astrophysik, Technische Universit\"at Berlin, Hardenbergstra{\ss}e 36, 10623 Berlin, Germany
}

\date{Accepted XXX. Received YYY; in original form ZZZ}

\pubyear{2021}

\begin{document}
\label{firstpage}
\pagerange{\pageref{firstpage}--\pageref{lastpage}}
\maketitle

\begin{abstract}
%This is a simple template for authors to write new MNRAS papers.
%The abstract should briefly describe the aims, methods, and main results of the paper.
%It should be a single paragraph not more than 250 words (200 words for Letters).
%No references should appear in the abstract.
The Rayleigh--Taylor (RT) instability is omnipresent in the physics of inversely density-stratified fluids subject to effective gravitational acceleration. In astrophysics, a steep stratification of the ambient medium can fragment a bubble shell faster due to a strongly time-dependent RT instability, causing the classical constant gravity models to fail. We derive the time-dependent instability criteria analytically for the cases of constant, exponential, and power-law accelerations, verifying them through high-resolution numerical simulations. Our results show that (1) even in the linear phase there is a term opposing exponential growth, (2) non-linear growth approaches asymptotically the solution found by Fermi and von Neumann, (3) the interpenetrating spikes and bubbles promote a significant mixing, with the fractal dimension of the interface approaching 1.6, only limited by numerical diffusion, and (4) the probability density function (PDF) for the passive scalar to study mixing becomes increasingly sharper peaked for power-law and exponential acceleration. Applying our solutions to stellar wind bubbles, young supernova remnants (SNRs), and superbubbles (SBs), we find that the growth rate of the RT instability is generally higher in the shells of wind-blown bubbles in a power-law stratified medium than in those with power-law rising stellar mechanical luminosities, Tycho-like than Cas~A-like SNRs, and one-sided than symmetric SBs. The recently observed \textit{eROSITA} bubbles indicate smooth rim surfaces, implying that the outer shell has not been affected by RT instabilities. Therefore the dynamical evolution of the bubbles suggests maximum final ages that are significantly above their current age, which we estimate to be about $\unit[20]{Myr}$.
\end{abstract}


%%%%%%%%%%%%%%%%%%%%%%%%%%%%%%%%%%%%%%%%%%%%%%%%%%

%%%%%%%%%%%%%%%%% BODY OF PAPER %%%%%%%%%%%%%%%%%%

\section{Introduction}
\label{sec:intro}

The Rayleigh--Taylor (RT) instability \citep{Ray:83,Tay:50} occurs on the interface between two fluids of different densities when an acceleration is directed from the heavier to the lighter fluid, or, in general, when the density gradient is misaligned in the opposite direction to the pressure gradient, leading to a baroclinic generation of vorticity.

By tapping into the reservoir of free energy available to such a system, small interfacial disturbances first grow exponentially, but shortly afterwards saturate and develop a characteristic mushroom-like shape, at least for some time. Then, larger wavelengths (which have yet to reach saturation) take over, with the flow eventually exhibiting self-similar behaviour as a result of non-linear mode interaction and successive wavelength saturation. This culminates in the formation of a turbulent mixing zone between the two fluids \citep{You:84}.

In its original form, the RT instability is driven by the gravitational field of a medium stratified in such a way that the total potential energy is not in a minimum. Many examples including, amongst others, surface tension and magnetic fields are treated in the classical textbook of \cite{Cha:61}. 

The problem areas RT instabilities are encountered in science and technology are manifold, and reach from the hydrodynamic mixing of fluids \citep{Sha:84} to inertial confinement fusion \citep{Nak:96}. In astrophysics, the RT instability is ubiquitous, since contact surfaces, separating different gases, occur naturally when sources of matter, momentum, and energy interact with the surrounding medium. If the acceleration that the interface thus experiences is in the same direction as the density contrast across it, an RT instability is inevitable, as every acceleration is equivalent to an oppositely directed gravitational field in the rest frame of the interface. Prominent examples are stellar wind bubbles \citep{Wea:77}, young supernova remnants \citep[SNRs;][]{Gul:73,Gul:75,Shi:78}, superbubbles \citep[SBs;][]{Low:88,Low:89}, accretion discs \citep{Wan:83}, etc. The general assumption is that the acceleration or deceleration of one fluid with respect to the other is constant. While this frequently holds, there are also many counterexamples, in particular when the contact surface meets a steep density or pressure gradient, causing a substantial rate of change in the velocity. However, so far this has not been treated in a systematic manner, which is the purpose of the present paper.

In order to assess the importance of the time-dependent RT instability, one has to compare the dynamical time-scale with the jerk time-scale, the latter being the rate of change of the acceleration. If the jerk time-scale is the shorter one, the analysis, which we are going to present here, should be applied. We are going to consider a fairly general class of time-dependent accelerations, ranging from an exponential to a power-law dependence. Finally, this allows us also to treat the so-called \textit{eROSITA} bubbles (EBs), spheroidal soft-X-ray-emitting structures that emanate from the Galactic Centre (GC) into the density-stratified halo with extensions of about $\unit[14]{kpc}$ on both sides \citep{Pre:20}. As such, they generously enclose the earlier discovered gamma-ray-emitting `\textit{Fermi} bubbles' \citep[FBs;][]{Dob:10,Su:10,Ack:14} and are possibly causally connected to them, with the FBs driving the expansion of the EBs and both objects being associated with the same (gradual or instantaneous) energy release in the GC region \citep{Pre:20}. Like the FBs, the EBs show a smooth structure, notably at the top and bottom, where the expansion velocity is highest, and the RT instability is supposed to happen most violently, thus leading to the fragmentation of the shell. Since this has not been observed yet, we are able to infer an upper hydrodynamical limit for their lifetimes.

The paper is structured as follows: in Sec.~\ref{sec:anasol}, analytical descriptions are presented for both the linear and the non-linear regime of the time-dependent RT instability, with the former including a derivation of a generalized dispersion relation via linear perturbation theory. Section~\ref{sec:numsol} is dedicated to the numerical analysis of the two-dimensional evolution of a single-mode RT instability for exponential and power-law variations of the acceleration. Section~\ref{sec:isbubbles} is devoted to astrophysical applications, notably to interstellar shells and supershells, and Sec.~\ref{sec:concl} closes the paper with our conclusions.

%%%%%%%%%%%%%%%%%%%%%%%%%%%%%%%%%%%%%%%%%%%%%%%%%%
\section{Analytical treatment}
\label{sec:anasol}

\begin{figure}
\centering
\includegraphics[width=\columnwidth]{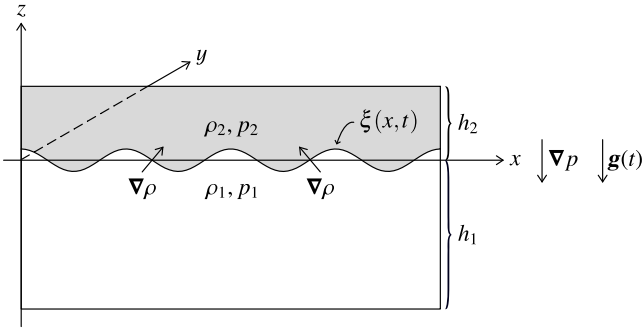}
\caption{Possible setup for the time-dependent Rayleigh--Taylor instability (for details see Section~\ref{sec:anasol}).}
\label{im:setup}
\end{figure}

We consider two distinct perfect fluids of density $\rho_1$ and $\rho_2$ under the influence of a time-dependent effective gravitational field $\mathbfit{g}(t)$, with the fluid of density $\rho_2$ resting on top. We emphasize that $\mathbfit{g}(t)$ can also be associated with the time-dependent acceleration or deceleration of an entire region. To keep the geometry simple, we use a Cartesian coordinate system in which the $z$-axis is pointing vertically upwards and the $xy$-plane is the equilibrium surface separating the two fluids, which are taken to be at rest initially. The acceleration vector is defined to be positive if it points from the upper to the lower fluid, that is $\mathbfit{g}(t)=-g(t)\,\hat{\mathbfit{e}}_z$. We further assume that both fluid layers have finite heights, namely $h_1$ for the bottom fluid and $h_2$ for the top one. The interface between the two fluids is slightly distorted, with $\xi(x,t)$ denoting the displacement of the surface in the $z$-direction at position $x$ and time $t$ (i.e.~we only consider perturbations that are independent of the $y$-coordinate; see Fig.~\ref{im:setup}).

%%%%%%%%%%%%%%%%%%%%%%%%%%%%%%%%%%%%%%%%%%%%%%%%%%
\subsection{Linear regime}
\label{sec:linreg}
During the early phase of the instability, when the amplitude of the perturbation is still small in comparison to its wavelength, we can neglect the non-linear term, $(\mathbfit{u}\cdot\nabla\mathbfit)\mathbfit{u}$, in the equation of fluid motion, with $\mathbfit{u}$ denoting the flow velocity. This implies that we have a potential flow with $\mathbfit{u}=-\nabla\varphi$, where $\varphi$ is the velocity potential. If we further assume that the fluid is incompressible, its dynamics is completely governed by
\begin{equation}
\nabla\left(-\pde{\varphi}{t}+\frac{P}{\rho}+\Phi\right)=\bmath{0}\,,
\label{eq:euler}
\end{equation}
where $P$ is the fluid pressure and $\Phi$ the potential for irrotational body forces. The integration of Eq.~(\ref{eq:euler}) is straightforward and yields
\begin{equation}
-\pde{\varphi}{t}+\frac{P}{\rho}+\Phi=f(t)\,.
\end{equation}
We note that $f(t)$ can be set to zero without any loss of generality as one can always add to $\varphi$ an arbitrary function of time without changing the relation $\mathbfit{u}=-\nabla\varphi$ \citep[cf.][]{Pad:00}. Hence we have
\begin{equation}
 -\pde{\varphi}{t}+\frac{P}{\rho}+\int_0^z g(t)\,dz'=0\,,
\end{equation}
or, equivalently,
\begin{equation}
P = -\rho \int_0^z g(t)\, dz'+\rho\,\pde{\varphi}{t}\,.
\label{eq:pexpr}
\end{equation}
Since the interface between the two fluids is physically represented by a contact discontinuity, the pressure has to be continuous across the surface $\xi(x,t)$, that is $P_1=P_2$. Therefore,
\begin{equation}
\begin{split}
-\rho_1 \, &\int_0^\xi g(t)\, dz+\rho_1\,  \left.\pde{\varphi}{t}\right|_1\\
&= -\rho_2\, \int_0^\xi g(t)\, dz+\rho_2\,  \left.\pde{\varphi}{t}\right|_2 \quad \textrm{at } z=\xi(x,t)\,,
\end{split}
\end{equation}
which, since $g(t)$ does not change appreciably over $\xi$, reduces to
\begin{equation}
\begin{split}
 -\rho_1\, &g(t)\, \xi +\rho_1 \, \left.\pde{\varphi}{t}\right|_1\\ 
&= -\rho_2\, g(t)\, \xi +\rho_2 \, \left.\pde{\varphi}{t}\right|_2 \quad \textrm{at } z=\xi(x,t)\,,
\end{split}
\end{equation}
yielding
\begin{equation}
\xi = \frac{1}{g(t)\, (\rho_1-\rho_2)} \, \left(\rho_1 \, \left.\pde{\varphi}{t}\right|_1- 
\rho_2 \, \left.\pde{\varphi}{t}\right|_2\right)\,.
\label{eq:xi}
\end{equation}
By the same token, the velocity in $z$-direction, $u_z$, has to be continuous across the interface, that is,
\begin{equation}
u_z= -\left.\pde{\varphi}{z} \right|_1 = -\left.\pde{\varphi}{z} \right|_2 \quad \text{at } z=\xi(x,t)\,.
\end{equation}
To the lowest order, $u_z$ is also equal to the rate of change of the displacement of the interface. Hence we can write $u_z=-\partial \varphi/\partial z=\partial \xi/\partial t$. Substituting for $\xi$ the expression given in Eq.~(\ref{eq:xi}) yields
\begin{equation}
\begin{split}
\pde{\varphi}{z} &=
-\frac{\dot{g}}{g^2\,(\rho_1-\rho_2)} \,
\left(\rho_2 \, \left.\pde{\varphi}{t}\right|_2- 
\rho_1 \, \left.\pde{\varphi}{t}\right|_1 \right)\\
&\quad+\frac{1}{g\, (\rho_1-\rho_2)} \, \left(\rho_2 \, \left.\ppde{\varphi}{t}\right|_2- 
\rho_1 \, \left.\ppde{\varphi}{t}\right|_1\right)\,,
\end{split}
\end{equation}
or, equivalently,
\begin{equation}
\begin{split}
g\,(\rho_1-\rho_2) \, \pde{\varphi}{z} &=
\rho_2\, \left.\ppde{\varphi}{t}\right|_2 -
\rho_1\, \left. \ppde{\varphi}{t}\right|_1\\
&\quad-\frac{\dot{g}}{g} \, \left( \rho_2\,\left. \pde{\varphi}{t}\right|_2 -
\rho_1\,\left. \pde{\varphi}{t}\right|_1\right)\,,
\end{split}
\label{eq:pressurebalance}
\end{equation}
with dots denoting time derivatives.

We now have to determine the form of $\varphi_1$ and $\varphi_2$. For the incompressible fluid we are studying, the continuity equation reduces to $\nabla\cdot\mathbfit{u}=0$. This implies that $\varphi$ must satisfy Laplace's equation, $\nabla^2\varphi=0$, and is independent of the $y$-coordinate (because of our assumption above). If we substitute the linear wave ansatz $\varphi=f(z)\,\exp[\i\,(k\,x-\omega\,t)]$ ($k$ is the perturbation wave number and $\omega$ the angular frequency) into Laplace's equation we find that
\begin{equation}
f(z)=C_1\,\exp(k\,z)+C_2\,\exp(-k\,z)\,,
\end{equation}
with $C_1$ and $C_2$ being integration constants. At the top ($z=h_2$) and bottom ($z=-h_1$) surface the velocity field has to vanish since the fluid layers end there (and there is nothing to perturb). Consequently (primes indicate derivatives with respect to the argument),
\begin{equation}
\begin{split}
\left.\pde{\varphi}{z}\right|_{h_2} &=f'(h_2) \, \exp[\i\,(k\,x-\omega\, t)]\\
&=k\,[C_1\,\exp(k\,h_2)-C_2\,\exp(-k\,h_2)]\\
&\quad\times\exp[\i\,(k\,x-\omega\,t)]=0
\end{split}
\label{eq:intconst1}
\end{equation}
and
\begin{equation}
\begin{split}
\left.\pde{\varphi}{z}\right|_{-h_1} &=f'(-h_1) \, \exp[\i\,(k\,x-\omega\, t)]\\
&=k\,[C_1\,\exp(-k\,h_1)-C_2\,\exp(k\,h_1)]\\
&\quad\times\exp[\i\,(k\,x-\omega\,t)]=0\,,
\end{split}
\label{eq:intconst2}
\end{equation}
which can be used to express one integration constant in terms of the other. From Eq.~(\ref{eq:intconst1}), $C_2=C_1\,\exp(2\,k\,h_2)$, and from Eq.~(\ref{eq:intconst2}), $C_2=C_1\,\exp(-2\,k\,h_1)$, and thus the solutions in the respective layers are
\begin{equation}
\varphi_1=A_1\,\cosh[k\,(z+h_1)]\,\exp[\i\,(k\,x-\omega\,t)]]
\end{equation}
and
\begin{equation}
\varphi_2=A_2\,\cosh[k\,(z-h_2)]\,\exp[\i\,(k\,x-\omega\,t)]]\,,
\end{equation}
with $A_1=2\,C_1\,\exp(-k\,h_1)$ and $A_2=2\,C_1\,\exp(k\,h_2)$. Inserting both solutions into the pressure balance equation (\ref{eq:pressurebalance}) we get two conditions valid at the interface, which can be rearranged to express the ratio $A_2/A_1$. For the lower fluid layer (index 1) we have
\begin{equation}
\begin{split}
g \, &(\rho_1-\rho_2) \, A_1 \, k \, \sinh\left[k\,(z+h_1)\right]\, \exp[\i\,(k\,x-\omega\,t)]\\
&=\left(\omega^2-\i\,\omega\, \frac{\dot{g}}{g}\right)
\, \Bigg\{\rho_1\, A_1 \, \cosh[k\, (z+h_1)]\\
&\quad-\rho_2\,A_2\, \cosh[k\,(z-h_2)]\Bigg\}\, \exp[\i\,(k\,x-\omega\,t)]\,,
\end{split}
\end{equation}
leading to
\begin{equation}
\begin{split}
\frac{A_2}{A_1} &= 
\left\{-\rho_2\, \left(\omega^2-\i\,\omega \,\frac{\dot{g}}{g}\right)\, \cosh[k\,(z-h_2)]\right\}^{-1}\\
&\quad \times \Bigg\{ g\, (\rho_1-\rho_2)\,k \, \sinh[k\,(z+h_1)]\\
&\quad -\rho_1\,\left(\omega^2-\i\,\omega\, \frac{\dot{g}}{g}\right)\, \cosh[k\,(z+h_1)]\Bigg\}\,.
\end{split}
\end{equation}
For the upper fluid layer (index 2) we obtain
\begin{equation}
\begin{split}
g\,&(\rho_1-\rho_2)\, A_2\,k \, \sinh[k\, (z-h_2)] \, \exp[\i \,(k\, x -\omega\, t)]\\
&= \left(\omega^2-\i \, \omega \, \frac{\dot{g}}{g}\right)\,
\bigg\{\rho_1\, A_1\, \cosh[k\,(z+h_1)]\\
&\quad -\rho_2\, A_2\,\cosh[k\,(z-h_2)]\bigg\}\, \exp[\i \, (k\, x- \omega\, t)]\,,
\end{split}
\end{equation}
and therefore
\begin{equation}
\begin{split}
\frac{A_2}{A_1} &=
\Bigg\{g \, (\rho_1-\rho_2)\, k \, \sinh[k\,(z-h_2)]\\
&\quad +\rho_2\,\left(\omega^2-\i\,\omega\, \frac{\dot{g}}{g}\right) \, \cosh[k\,(z-h_2)]\Bigg\}^{-1}\\
&\quad \times \left\{\rho_1 \,\left(\omega^2-\i\,\omega\, \frac{\dot{g}}{g}\right)\, \cosh[k\,(z+h_1)]\right\}\,.
\end{split}
\end{equation}
We can now equate both expressions for $A_2/A_1$, which yields, after a little algebra,
\begin{equation}
\omega^2-\i \, \omega \, \frac{\dot{g}}{g}  = \frac{k \, g\, (\rho_1-\rho_2)}{
\rho_1\, \coth[k\,(z+h_1)]+\rho_2 \,\coth[k\,(h_2-z)]}\,.
\end{equation}
This amounts to solving the quadratic equation
\begin{equation}
B\,\omega^2-\i\,B\,\frac{\dot{g}}{g}\,\omega - k\,g\, (\rho_1-\rho_2) =0\,,
\end{equation}
where $ B=\rho_1\, \coth(k\,h_1)+\rho_2 \,\coth(k\,h_2)>0$,
by observing that $z \to 0$ at the interface. The solutions of this dispersion relation are given by
\begin{equation}
\omega = \frac{\i}{2\, g} \, \left[ \dot{g}\pm \sqrt{\dot{g}^2 + \frac{4\,k\,g^3}{B}\,(\rho_2-\rho_1)}\right]\,.
\label{eq:sol_disprel}
\end{equation}

An unstable configuration emerges if $\im(\omega)>0$, which is satisfied by the following parameter regimes:
\begin{enumerate}
\item \tabto{1cm}$g>0$, $\dot{g}>0$,    and $(\rho_2-\rho_1)\ge -\frac{B\,\dot{g}^2}{4\,k\,g^3}$,
\item \tabto{1cm}$g>0$, $\dot{g}\le 0$, and $(\rho_2-\rho_1)>0$,
\item \tabto{1cm}$g<0$, $\dot{g}<0$,    and $(\rho_2-\rho_1)\le -\frac{B\,\dot{g}^2}{4\,k\,g^3}$,
\item \tabto{1cm}$g<0$, $\dot{g}\ge 0$, and $(\rho_2-\rho_1)<0$.
\end{enumerate}
We note that the cases (ii) and (iv) include the classical time-independent RT instability criterion, $\mathbfit{g}\cdot \nabla\rho<0$ with $\dot{\mathbfit{g}}=\bmath{0}$.

There are three limiting cases for the growth rate $\sigma=|\im(\omega)|$. First, for $k\,h_1\gg 1$ and $k\,h_2\gg 1$ (deep fluid layers and/or short waves) it holds that
\begin{equation}
\sigma\simeq\left|\frac{\dot{g}}{2\,g}\pm\sqrt{\frac{\dot{g}^2}{4\,g^2}+k\,g\,\mathscr{A}}\right|\,,
\label{eq:growthrate_limit_A}
\end{equation}
with $\mathscr{A}=(\rho_2-\rho_1)/(\rho_2+\rho_1)$ being the (dimensionless) Atwood number, which characterizes the strength of the stratification and ranges from $-1$ to $1$. The magnitude of $\mathscr{A}$ generally affects the RT instability not only quantitatively but also qualitatively. For $|\mathscr{A}|\lesssim 0.1$, the mixing zone formed by the penetration of the heavier into the lighter fluid (as `spikes') and the lighter into the heavier one (as `bubbles') expands symmetrically away from the initial position of the density interface. Higher values of $|\mathscr{A}|$ break this symmetry, with the heavier spikes penetrating deeper than the lighter bubbles, which is due to the narrowing of the spikes, and a consequent reduction of drag on their heads \citep{And:10}. If, in particular, $k\gg 1$, the second term in the discriminant of Eq.~(\ref{eq:growthrate_limit_A}) dominates. Then, the growth rate becomes independent of $\dot{g}$ and increases with increasing $k$ ($g$ is always bounded due to physical limits),
\begin{equation}
\sigma\simeq \sqrt{k\,g\,\mathscr{A}}\,.
\end{equation}
Second, for $k\,h_1\ll 1$ and $k\,h_2\ll 1$ (shallow fluid layers and/or long waves) we find
\begin{equation}
\sigma\simeq\left|\frac{\dot{g}}{2\,g}\pm\sqrt{\frac{\dot{g}^2}{4\,g^2}+k^2\,\frac{g\,(\rho_2-\rho_1)\,h_1\,h_2}{\rho_2\,h_1+\rho_1\,h_2}}\right|\,.
\label{eq:growthrate_limit_B}
\end{equation}
If, in particular, $k\ll 1$, then, by taking the positive root in Eq.~(\ref{eq:growthrate_limit_B}), $\sigma\simeq |\dot{g}/g|$. And, finally, for $k\,h_1\gtrsim 1$ and $k\,h_2\ll 1$ we obtain
\begin{equation}
\sigma\simeq\left|\frac{\dot{g}}{2\,g}\pm\sqrt{\frac{\dot{g}^2}{4\,g^2}+k^2\,\frac{g\,(\rho_2-\rho_1)\,h_2}{\rho_2}}\right|\,.
\end{equation}

For the sake of deriving an equation that governs the dynamics of the perturbed interface, we write down the hydrostatic pressures on each side of the interface,
\begin{align}
P_1&=P_0-\rho_1\,g\,\xi\,,\\
P_2&=P_0-\rho_2\,g\,\xi\,,
\end{align}
with $P_0$ being the pressure at the interface when it was still unperturbed along $z=0$. With the acceleration pointing into the $(-z)$-direction, the pressure gradient must build up in the same direction, so that $P_1>P_2$, and $\rho_2>\rho_1$ for an unstable situation. Here, again, we assume that $g(t)$ does not change appreciably across the interface. Hence, according to Newton's second law,
\begin{equation}
\Delta P = P_1-P_2= (\rho_2-\rho_1)\,g\,\xi =\frac{\Delta F}{S}=\frac{m}{S}\,\ppde{\xi}{t}\,,
\label{eq:deltap}
\end{equation}
with $\Delta F$ being the pressure force difference across the surface $S$ at which the perturbation acts; $m$ in turn is the mass of the perturbed fluid column of height $H$, that is
\begin{equation}
m=m_1+m_2=(\rho_1+\rho_2)\,S\,H\,.
\label{eq:m1plusm2}
\end{equation}
Since the wave perturbation decays away from the contact surface like $\exp(-k\,|z|)$, a typical distance can be defined via $k\,z\sim k\,H\sim 1$, and therefore $H\sim 1/k$. Substituting this together with Eq.~(\ref{eq:m1plusm2}) into Eq.~(\ref{eq:deltap}) we arrive at the differential equation of the surface motion,
\begin{equation}
(\rho_2-\rho_1)\,g\,\xi =\frac{\rho_1+\rho_2}{k}\,\ppde{\xi}{t}\,,
\end{equation}
or, equivalently, in terms of the perturbation amplitude $\eta=\eta(t)$,
\begin{equation}
\ddot{\eta}(t)=\frac{k\,g(t)}{B}\,(\rho_2-\rho_1)\,\eta(t)\,,
\label{eq:surfaceeq}
\end{equation}
with the last expression also taking into account finite fluid layer heights \citep{For:09}.

Solving Eq.~(\ref{eq:surfaceeq}) for the classical time-independent case,
\begin{equation}
g(t)=g_0=\mathrm{const.}\,,
\label{eq:mod_con}
\end{equation}
together with the initial conditions $\eta(t=0)=\eta_0$ and $\dot{\eta}(t=0)=\varv_0$, leads to
\begin{equation}
\begin{split}
\eta(t) &= \frac{\eta_0}{2}\,\left[\exp(\sigma_0\,t)+\exp(-\sigma_0\,t)\right]\\
&\quad+\frac{\varv_0}{2\,\sigma_0}\,\left[\exp(\sigma_0\,t)-\exp(-\sigma_0\,t)\right]\\
&=\eta_0\,\cosh(\sigma_0\,t)+\frac{\varv_0}{\sigma_0}\,\sinh(\sigma_0\,t)\,,
\label{eq:eta_con}
\end{split}
\end{equation}
where $\sigma_0=\sqrt{k\,g_0\,(\rho_2-\rho_1)/B}\propto \lambda^{-1/2}$ is the classical growth rate, and $\lambda=2\,\pi/k$ is the perturbation wavelength. Hence short-wavelength perturbations are the ones that grow fastest during the linear regime. Furthermore it is apparent from inspection of Eq.~(\ref{eq:eta_con}) that the early growth of the RT instability is not purely exponential, as is often wrongly assumed in the astrophysical literature, but is governed by a combination of stimulating and inhibiting terms. As a result, calculations that neglect the inhibiting terms strongly overestimate the linear amplitude of the RT instability \citep{Liu:20} and thus underestimate its growth time!

Following \cite{Kul:91}, the solution of Eq.~(\ref{eq:surfaceeq}) with an exponential acceleration law of the form
\begin{equation}
g(t)=g_0\,\exp\left(\delta\,\frac{t}{\tau}\right)
\label{eq:mod_exp}
\end{equation}
can be found via the variable substitution $\psi=\psi_0\,\exp[\delta\,t/(2\,\tau)]$ with the initial value $\psi_0=2\,\sigma_0\,\tau/|\delta|$, and $\psi$ lying in the interval $[0,\psi_0]$ for $\delta <0$ and in the interval $[\psi_0,\infty]$ for $\delta >0$. Equation (\ref{eq:surfaceeq}) then becomes the modified Bessel equation,
\begin{equation}
\psi^2\,\eta''(\psi)+\psi\,\eta'(\psi)-\psi^2\,\eta(\psi)=0\,.
\end{equation}
Solving this differential equation with the same initial conditions as before, which transform to $\eta(\psi=\psi_0)=\eta_0$ and $\eta'(\psi=\psi_0)=\varv_0/\sigma_0$, we get
\begin{equation}
\begin{split}
\eta(\psi)&=\psi_0\,\{\eta_0\,[K_1(\psi_0)\,I_0(\psi)+I_1(\psi_0)\,K_0(\psi)]\\
&\quad+\frac{\varv_0}{\sigma_0}\,|K_0(\psi_0)\,I_0(\psi)-I_0(\psi_0)\,K_0(\psi)|\}\,,
\end{split}
\end{equation}
where $I_\nu(\psi)$ and $K_\nu(\psi)$ are the modified Bessel functions of the first and second kind, respectively.

For accelerations that instead obey a power law of the form
\begin{equation}
g(t)=g_0\,\left(\frac{t}{\tau}\right)^\delta+g_1\,,
\label{eq:mod_pow}
\end{equation}
with an arbitrary non-zero exponent $\delta$, Eq.~(\ref{eq:surfaceeq}) has to be solved numerically. We find that analytical integration is possible, though, for the special case of $g_1=0$ and $\delta=n$ with $n=1,2,3,\dots$. By rescaling the time variable as 
\begin{equation}
\psi=\frac{\sigma_0^2}{(n+2)^2\,\tau^n}\,t^{n+2}\,,
\end{equation}
Eq.~(\ref{eq:surfaceeq}) expands to the confluent hypergeometric limit equation,
\begin{equation}
\psi\,\eta''(\psi)+p\,\eta'(\psi)-\eta(\psi)=0\,, \quad \text{with } p=\frac{n+1}{n+2}\,.
\end{equation}
This has the general solution
\begin{equation}
\eta(\psi)=C_1\,{}_0F_1(;p;\psi)+ C_2\,(-\psi)^{1-p}\,{}_0F_1(;2-p;\psi)\,,
\end{equation}
where $C_1$ and $C_2$ are constants of integration, and ${}_0F_1(;p;\psi)$ is the confluent hypergeometric limit function, which is related to the modified Bessel function of the first kind via \citep[see e.g.][]{Abr:64}
\begin{equation}
{}_0F_1(;p;\psi) = (\sqrt{\psi})^{1-p}\,I_{p-1}(2\,\sqrt{\psi})\,\Gamma(p)\,,
\end{equation}
with $\Gamma(p)$ denoting the gamma function. Applying the usual initial conditions, which now read $\eta\rightarrow \eta_0$ and $\eta'\rightarrow \varv_0\,\{\tau^n/[(n+2)^n\,\sigma_0^2\,\psi^{n+1}]\}^{1/(n+2)}$ as $\psi\rightarrow 0$, we end up with
\begin{equation}
\begin{split}
\eta(\psi)&=\eta_0\,{}_0F_1\left(;\frac{n+1}{n+2};\psi\right)\\
&\quad+\varv_0\,\left[\frac{(n+2)^2\,\tau^n\,\psi}{\sigma_0^2}\right]^{1/(n+2)}\,{}_0F_1\left(;\frac{n+3}{n+2};\psi\right)\,.
\end{split}
\end{equation}

As soon as the perturbation amplitude reaches a size of order $\lambda/2$, its exponential growth slows down and the non-linear phase begins \citep{You:84}.

%%%%%%%%%%%%%%%%%%%%%%%%%%%%%%%%%%%%%%%%%%%%%%%%%%
\subsection{Non-linear regime}
\label{sec:nonlinreg}
As time goes by, the growth of the undulations at the contact discontinuity eventually enters the so-called `self-similar phase', which was first quantitatively analyzed by \cite{Fer:53}. The self-similar growth of the RT-unstable structures is governed by the relation \citep[see e.g.][]{Ris:04}
\begin{equation}
\dot{\eta}(t)=2\,\sqrt{\frac{\alpha\,g(t)}{B}\,(\rho_2-\rho_1)\,\eta(t)}\,.
\label{eq:fermineumann}
\end{equation}
The dimensionless parameter $\alpha$ may be thought of as a measure of the efficiency of potential energy release. Experiments and simulations suggest that $\alpha$ lies in the range of 0.02 to 0.1 \citep{Wei:12}, with the particular value depending on Atwood number, initial conditions, evolution time, and dimensionality \citep{You:84}. Equation (\ref{eq:fermineumann}) can be obtained via a self-similarity assumption \citep{Ris:04} or from an energy argument \citep{Coo:04}. While the former entails a rigorous derivation from first principles using the Navier-Stokes equations, the latter recognizes that $\dot{\eta}$ is proportional to the net mass flux through the interface and models the vertical velocity fluctuations at the interface through a generalization of the terminal velocity equation for a falling sphere with a diameter proportional to $\eta$ \citep{Coo:09}. For constant acceleration (Eq.~\ref{eq:mod_con}), density contrast, $B$, and $\alpha$, the solution to Eq.~(\ref{eq:fermineumann}) is (taking only the positive root as physically realizable)
\begin{equation}
\begin{split}
\eta(t) &=\frac{\alpha\,g_0}{B}\,(\rho_2-\rho_1)\,(t-t_0)^2\\
&\quad+2\,\sqrt{\frac{\alpha\,g_0}{B}\,(\rho_2-\rho_1)\,\eta_0}\,(t-t_0)+\eta_0\,.
\label{eq:fncon}
\end{split}
\end{equation}
Here, $\eta_0=\eta(t_0)$ can represent either a virtual starting amplitude, that effectively depends on how long it takes the flow to become self-similar, which in turn depends on the spectrum of the initial perturbations, or, alternatively, the perturbation amplitude at the moment when the RT instability first reaches the non-linear regime, provided that this happens at time $t=t_0$ \citep{Cab:06}.

Turning now to time-dependent accelerations, in particular those given by Eqs.~(\ref{eq:mod_exp}) and (\ref{eq:mod_pow}), we obtain the solutions
\begin{equation}
\begin{split}
\eta(t) &= \frac{4\,\alpha\,g_0\,\tau^2}{\delta^2\,B}\,(\rho_2-\rho_1)\,\left[\exp\left(\delta\,\frac{t}{2\,\tau}\right)-\exp\left(\delta\,\frac{t_0}{2\,\tau}\right)\right]^2\\
&\quad+\frac{4\,\tau}{\delta}\,\sqrt{\frac{\alpha\,g_0}{B}\,(\rho_2-\rho_1)\,\eta_0}\\
&\quad\times\left[\exp\left(\delta\,\frac{t}{2\,\tau}\right)-\exp\left(\delta\,\frac{t_0}{2\,\tau}\right)\right]+\eta_0
\end{split}
\end{equation}
and
\begin{equation}
\begin{split}
\eta(t)&=\frac{\alpha\,g_1}{B}\,(\rho_2-\rho_1)\,\left[t\,{}_2F_1\left(-\frac{1}{2},\frac{1}{\delta};\frac{\delta+1}{\delta};-\frac{g_0\,t^\delta}{g_1\,\tau^\delta}\right)\right.\\
&\quad\left.-t_0\,{}_2F_1\left(-\frac{1}{2},\frac{1}{\delta};\frac{\delta+1}{\delta};-\frac{g_0\,t_0^\delta}{g_1\,\tau^\delta}\right)\right]^2\\
&\quad+2\,\sqrt{\frac{\alpha\,g_1}{B}\,(\rho_2-\rho_1)\,\eta_0}\\
&\quad\times\left[t\,{}_2F_1\left(-\frac{1}{2},\frac{1}{\delta};\frac{\delta+1}{\delta};-\frac{g_0\,t^\delta}{g_1\,\tau^\delta}\right)\right.\\
&\quad\left.-t_0\,{}_2F_1\left(-\frac{1}{2},\frac{1}{\delta};\frac{\delta+1}{\delta};-\frac{g_0\,t_0^\delta}{g_1\,\tau^\delta}\right)\right]+\eta_0\,,
\end{split}
\end{equation}
respectively. Again we have set $\eta_0=\eta(t_0)$, and ${}_2F_1(p,q;r;\psi)$ denotes a hypergeometric function. A solution that does not involve any transcendental functions emerges for $g(t)=K\,t^{\beta-2}$, namely
\begin{equation}
\begin{split}
\eta(t)&=\frac{4\,\alpha\,K}{\beta^2\,B}\,(\rho_2-\rho_1)\,\left(t^{\beta/2}-t_0^{\beta/2}\right)^2\\
&\quad+\frac{4}{\beta}\,\sqrt{\frac{\alpha\,K}{B}\,(\rho_2-\rho_1)\,\eta_0}\,\left(t^{\beta/2}-t_0^{\beta/2}\right)+\eta_0\,.
\label{eq:fnpoweasy}
\end{split}
\end{equation}
It is clear that in the asymptotic limit of long times the first right-hand side term of Eqs.~(\ref{eq:fncon})--(\ref{eq:fnpoweasy}) dominates, with $t_0$ being negligible with respect to $t$. 

As for Atwood number magnitudes above 0.1 the RT mixing becomes asymmetric, Eqs.~(\ref{eq:fncon})--(\ref{eq:fnpoweasy}) need modification to account for the different spike and bubble $\alpha$'s, with the former always being at least as large as the latter \citep[see][and references therein]{Ban:20}. 

Furthermore it is important to note that for Eqs.~(\ref{eq:fncon})--(\ref{eq:fnpoweasy}) to be applicable, $\eta$ must be much larger than the diffusion (and viscous) scale, because otherwise the fluid's diffusivity (and/or viscosity) will introduce additional length- and time-scales into the problem, thereby ruling out self-similarity. In addition, the initial perturbations must be band-limited, because otherwise there will be a competition between the linear and the non-linear growth law for long-wavelength perturbations. On the other hand, the sum of the spike and bubble amplitudes must remain smaller than the smallest spatial extent of the domain in which the RT instability is developing in order to avoid the interference of any boundary effect \citep{Coo:09}.

%%%%%%%%%%%%%%%%%%%%%%%%%%%%%%%%%%%%%%%%%%%%%%%%%%
\section{Numerical treatment}
\label{sec:numsol}
\subsection{General simulation setup}
We verify the analytical solution relations from the previous section through two-dimensional numerical simulations of a single-mode RT instability, where a perturbation with a fixed wavelength is applied to the initial interface. The simulations are carried out with the publicly available\footnote{\url{https://bitbucket.org/rteyssie/ramses}} massively parallel octree-based adaptive mesh refinement (AMR) code \mbox{\textsc{ramses}} \citep{Tey:02}, which allows for solving the discretized Euler equations in their conservative form by means of an unsplit second-order accurate Monotonic Upstream-centered Scheme for Conservation Laws \citep[MUSCL;][]{Lee:79} Godunov method for polytropic gases. Since the problem involves the precise tracing of contact discontinuities, not an approximate but an exact Riemann solver is applied \citep[see e.g.][]{Tor:09}, together with the monotonized central-difference slope limiter \citep{Lee:77} to make the scheme total variation diminishing \citep[TVD;][]{Har:83a}.

In order to allow the RT spikes and bubbles to grow as symmetrically as possible, and thus to enable a straightforward comparison with the analytical solution derived via linear theory, which does not differentiate between these two structures \citep[at least if no corrections accounting for the non-linear mode coupling are applied; see][]{Liu:20}, we choose an Atwood number of $\mathscr{A}=0.1$. In our setup, this translates to densities of the heavy and light fluid of $\rho_2=\unit[1.22]{g\,cm^{-3}}$ and $\rho_1=\unit[1]{g\,cm^{-3}}$, respectively. At $t=0$, the acceleration is taken to be $g_0=\unit[1]{cm\,s^{-2}}$ in the downward $(-z)$ direction. The computational domain is square, with side length $L=\unit[1]{cm}$. Its left and right boundaries are set to be periodic, the top and bottom ones are reflective. The initial interface separating the denser from the less dense fluid is centered vertically at $z_0=L/2$, with the densities in the lower and upper half taken to be $\rho_1$ and $\rho_2$, respectively. The initial pressure profile obeys the condition of hydrostatic equilibrium, that is
\begin{equation}
  P(z,t=0) = 
  \begin{cases}
    P_\textrm{t}+\rho_2\, g_0\, (L-z) & z > z_0\,, \\
    P_\textrm{t}+\rho_2\, g_0\, z_0 + \rho_1\, g_0\, (z_0-z) & z \le z_0\,,
  \end{cases}
\end{equation}
where $P_\textrm{t}=\unit[10^3]{dyn\,cm^{-2}}$ is the pressure at the top of the domain. Setting the adiabatic index $\gamma=5/3$, for a perfect monoatomic gas, this gives a sound speed at the interface in the less dense medium of $a_\textrm{0,max}\approx \unit[40.8]{cm\,s^{-1}}$. The resulting Courant-limited adaptive time steps (the Courant number is set to 0.8) are not too small to lead to unacceptable long computing times on a medium-sized cluster, even at rather high resolutions, while still keeping the fluid compressibility sufficiently low (see below).

\begin{figure}%[!htb]
\centering
\includegraphics[width=\columnwidth]{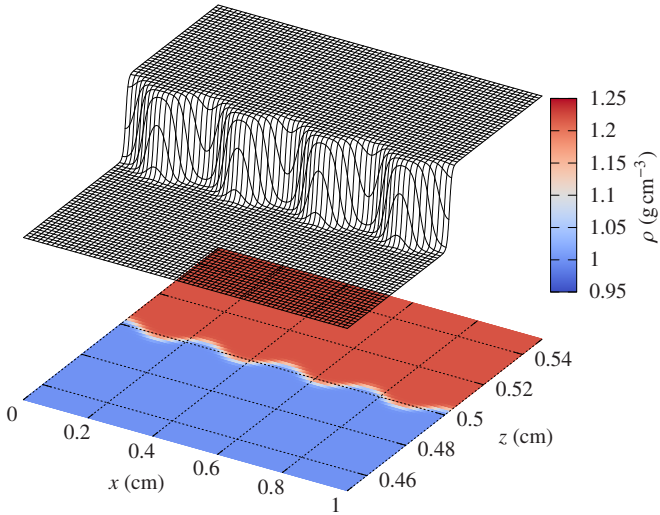}
\caption{Cutout of the initial density distribution applied in all simulations. The strength of the perturbation is exaggerated since not the same scale is used for the $x$- and $z$-axis.}
\label{im:ics}
\end{figure}

To initiate the instability, we slightly shift the density around the interface in a periodic fashion. We define the height of the perturbed interface at $t=0$ as
\begin{equation}
\xi_0(x)=\frac{\eta_0}{2}\,\left\{\cos\left[\frac{2\,\pi\,x}{\lambda}\right]+\cos\left[\frac{2\,\pi\,(L-x)}{\lambda}\right]\right\}+z_0\,,
\end{equation}
with the wavelength $\lambda=L/4$ and the amplitude $\eta_0=0.01\,\lambda$. The way the cosine part of the perturbation is calculated here prevents roundoff errors from introducing an asymmetry in the flow \citep[cf.][]{Alm:10}. The density is then perturbed as
\begin{equation}
\rho(x,z)=\rho_1+\frac{\rho_2-\rho_1}{2}\,\left\{1+\tanh\left[\frac{z-\xi_0(x)}{s}\right]\right\}\,,
\end{equation}
with the hyperbolic tangent profile providing a slight smearing of the initial interface over a smoothing length, $s$,  in order to avoid gridding errors. We set $s=0.005\,\lambda$ and enforce that the initial AMR grid close to the interface is always fully refined. The profile is plotted in Fig.~\ref{im:ics}.
\begin{figure*}
\resizebox{\hsize}{!}
{\includegraphics{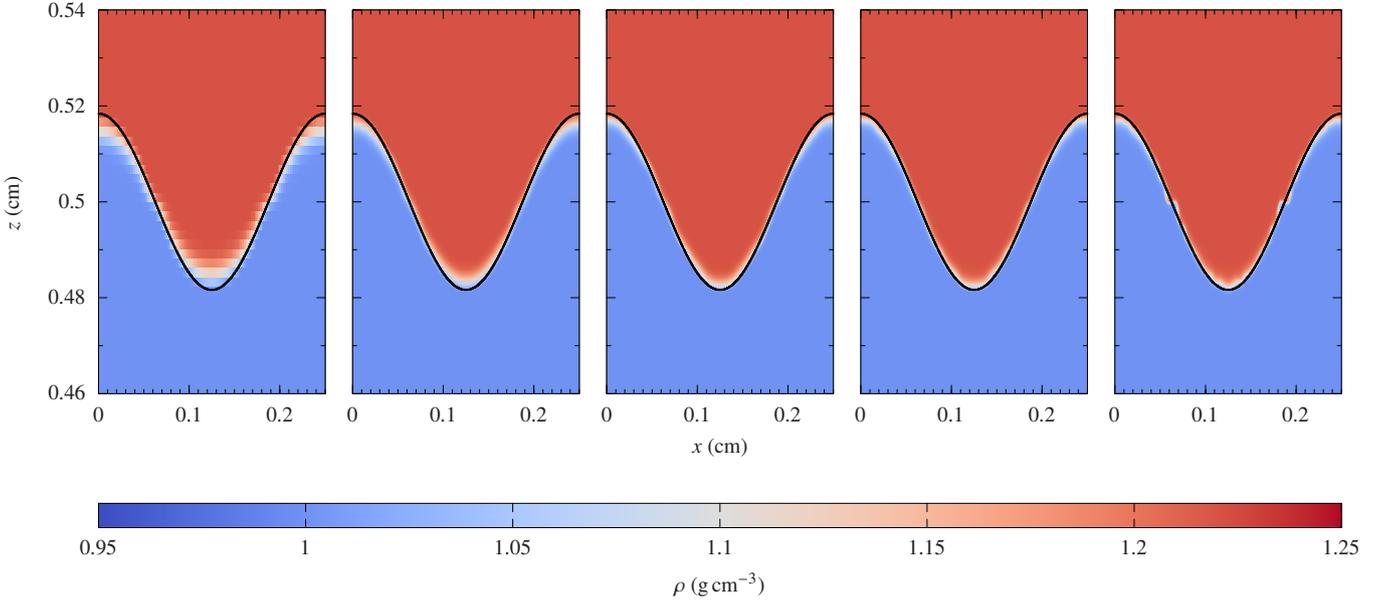}}
\caption{Single wavelength of the Rayleigh--Taylor instability under the influence of constant vertical acceleration of unity magnitude at time $t=\unit[1.7]{s}$, as calculated numerically with five different spatial resolutions. Shown is the density color-coded, with the analytical solution superimposed (black curve). The number of grid cells across the horizontal dimension ($x$) of the cut-out domain, from left to right, is 128, 256, 512, 1024, and 2048. Perturbations are exaggerated since not the same scale is used for the $x$- and $z$-axis.}
\label{im:resostudy}
\end{figure*}

Figure~\ref{im:resostudy} displays the results of a resolution study. Shown is a single wavelength of the RT instability at $t=\unit[1.7]{s}$ with the spatial resolution increasing from left to right. In all cases, the minimum refinement level of the Cartesian grid comprises $64^2$ cells in total, which corresponds to a basic spatial resolution of about $\unit[0.016]{cm}$. This grid is adaptively refined between three (leftmost panel) and seven (rightmost panel) levels in flow regions where the pressure and density gradients exceed 1 per cent of the local normalized values, with the spatial resolution doubling with each level. It can be seen that, as the effective resolution is increased, the numerical solution approaches the analytical one, but, simultaneously, more small scale structure develops, so that the calculation actually never strictly converges. The reason for this is the reduced amount of numerical dissipation present at higher resolutions that allows for secondary instabilities of Kelvin--Helmholtz \citep[KH;][]{Hel:68,Kel:71} type to develop \citep[cf.][]{Cal:02}. These arise from the shear flow between the interpenetrating RT spikes and bubbles. As a consequence, the RT instability can never operate completely free from the KH instability, whereas the reverse can absolutely occur in nature (e.g.~undulations of a water surface caused by blowing wind).

We find that, besides the spatial resolution, also the pressure of the medium controls the occurrence of these interface ripples, with higher pressure values having a suppressing effect. This is not surprising since increasing the pressure also increases the sound speed, which is the speed at which information is passed through the medium. Accordingly, a medium with a higher sound speed `becomes aware' of interface corrugations earlier and thus can counteract them more efficiently, provided that the speed of the perturbation remains sufficiently low -- the fluid behaves less compressible \citep[see e.g.][]{Lan:59}. The analytical solution, on the other hand, assumes a perfectly incompressible medium (with a sound speed that is formally infinite) and is therefore perfectly smooth at all times. In order to mimic this behaviour in our (compressional) numerical treatment as well as possible, we perform our fiducial simulations with an effective resolution of `only' $4096^2$ cells (cf.~the second panel from the right in Fig.~\ref{im:resostudy}), implying that structures with a size as small as $\unit[2.4\times 10^{-4}]{cm}$ can still be resolved. At that resolution, the contact discontinuity remains smooth for a sufficiently long time, while still matching the growth rate of the primary (RT) instability, as obtained from linear theory, correctly. 

We consider three different models: one with the vertical acceleration remaining constant over time (Eq.~\ref{eq:mod_con} with $g_0=\unit[1]{cm\,s^{-2}}$; model CON), which serves as a reference, and two with time-dependent accelerations, either obeying an exponential law (Eq.~\ref{eq:mod_exp} with $\tau=\unit[1]{s}$, $g_0=\unit[1]{cm\,s^{-2}}$, and $\delta=1/2$; model EXP) or a power law (Eq.~\ref{eq:mod_pow} with $\tau=\unit[1]{s}$, $g_0=g_1=\unit[1]{cm\,s^{-2}}$, and $\delta=4/5$; model POW). All models are evolved until $t=\unit[10]{s}$.

%%%%%%%%%%%%%%%%%%%%%%%%%%%%%%%%%%%%%%%%%%%%%%%%%%
\subsection{Model CON}
\label{sec:modcon}
\begin{figure*}
\resizebox{\hsize}{!}{\includegraphics{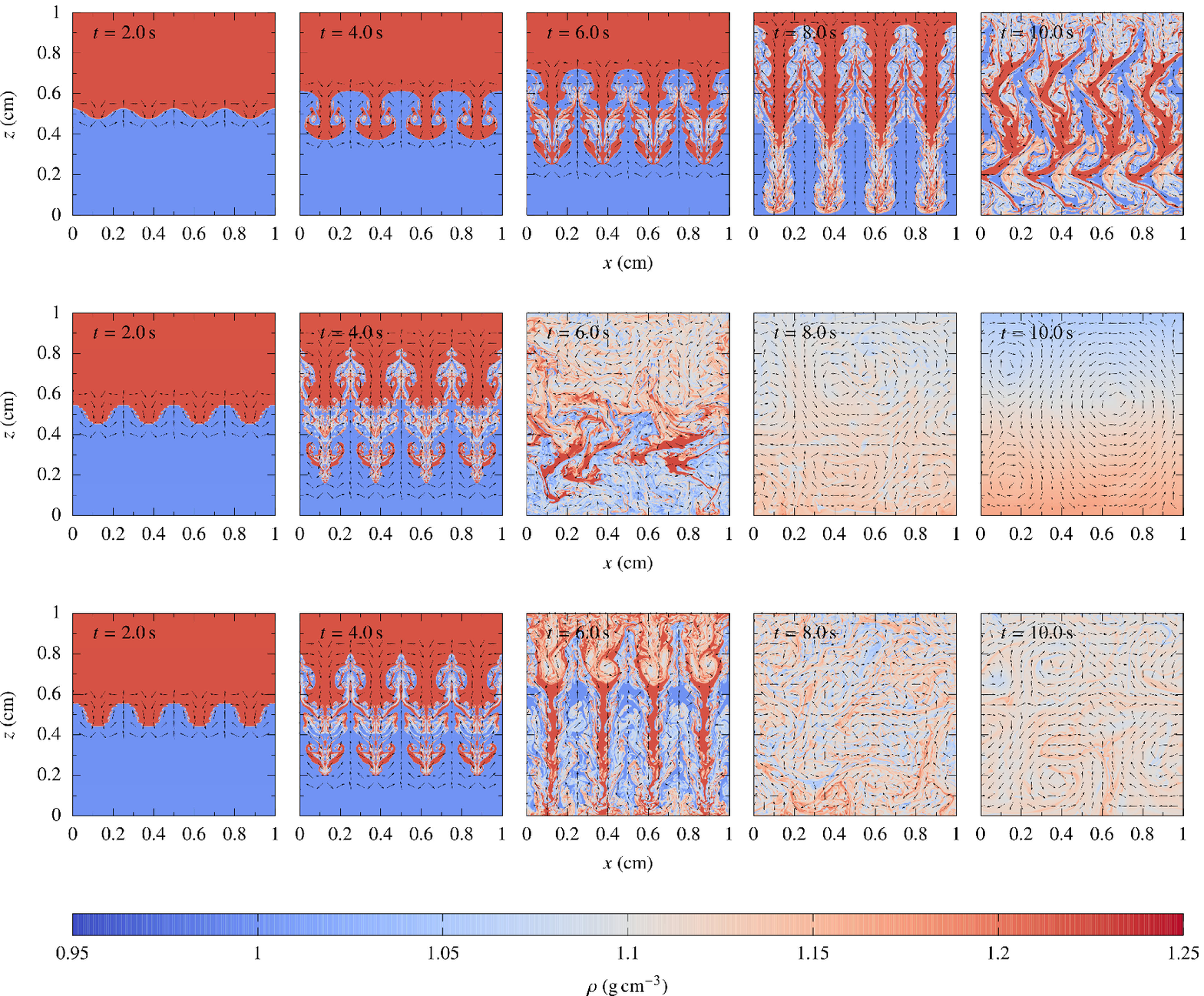}}
\caption{Temporal evolution of the Rayleigh--Taylor instability under the influence of constant (model CON; \emph{top row}), exponential (model EXP; \emph{middle row}), and power-law (model POW; \emph{bottom row}) vertical acceleration. Shown in each panel is the density color-coded, with the velocity field (arrows) superimposed. The magnitudes of the velocity vectors plotted range from $0.01$ to $\unit[1.00]{cm\,s^{-1}}$.}
\label{im:timeseries}
\end{figure*}
Density maps of the overall evolution, with the velocity field superimposed, are shown in the top-row panels of Fig.~\ref{im:timeseries}. The baroclinic torque that arises from the misalignment of the density and pressure gradients at the perturbed interface creates vorticity and induces a velocity field that increases the baroclinic torque, thus closing the self-exciting feedback loop \citep{Rob:16}. This torque, specified with regard to the axes of the rotating velocity vectors, is highest at the points where the angle the density and pressure gradient vectors enclose with each other is closest to $\pi/2$, namely for example roughly in the middle of the lateral outline of the RT spikes (and bubbles) in the snapshot at $t=\unit[2]{s}$, or `under' the `mushroom caps' in the snapshot at $t=\unit[4]{s}$. By the time, particularly after the instability has reached the reflecting boundaries of the domain ($t\sim \unit[8]{s}$), which drastically enhances the ongoing interaction and merging of the RT spikes and bubbles, the velocity field becomes more and more disordered, giving rise to increasingly irregular, chaotic structures -- the flow has developed into a turbulent state.
\begin{figure*}
\resizebox{\hsize}{!}{\includegraphics{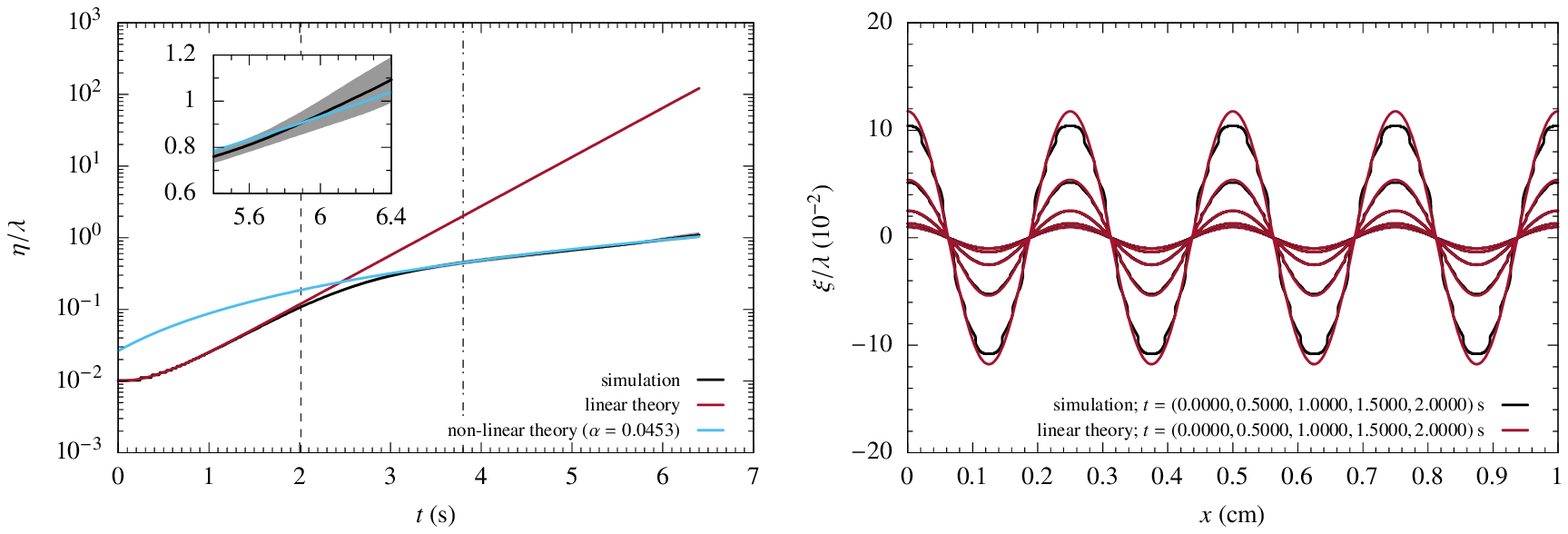}}
\caption{Results for model CON. \emph{Left panel}. The black solid line shows the numerically derived growth of the perturbation amplitude over time. This is the average of the (normalized) amplitudes of the Rayleigh--Taylor spikes and bubbles, which are represented by the upper and lower boundaries of the grey band, respectively, as can be seen in the inlay. The solution from linear theory is given by the red line, with the ending of this regime marked by the dashed vertical line. The blue line represents the evolution law for the non-linear regime. It is based on the rate coefficient $\alpha$, that is obtained via a least-squares fit to the part of the numerical solution lying to the right of the dash-dotted vertical line, which in turn marks the time at which the flow should have fully transitioned to non-linearity. \emph{Right panel}. The perturbed interface at five equally spaced times during the linear regime, calculated both numerically (black lines) and analytically (red lines).}
\label{im:rtanalysis_con}
\end{figure*}

A detailed comparison with the analytical solutions is presented in Fig.~\ref{im:rtanalysis_con}. The numerical solution curves, which are based on the actual shape of the contact discontinuity, $\xi(x,t)$, are produced with a `contact tracer' that scans through the grid, returning the cell positions in $z$-direction where $|\nabla\rho(x,z,t)|$ is maximum for given values of $x$ and $t$. The black line in the left plot denotes the mean penetration depth of the RT instability, $\eta(t)$, which we calculate from taking the arithmetic mean of the amplitudes of the RT spikes, $\eta_\text{s}(t)=|\min_x\xi(x,t)|$, and bubbles, $\eta_\text{b}(t)=|\max_x\xi(x,t)|$. These amplitudes are shown in the graph as well, namely as the upper and lower boundaries of the grey band, respectively, that encloses the black line (as seen in the inlay magnification). The narrowness of this band is a proof for the almost symmetric growth of these two structures under the conditions imposed.

We define the end of the linear regime through the time at which the relative error between the numerical and the analytical solution, as computed from Eq.~(\ref{eq:eta_con}), exceeds 10 per cent for the first time, which is marked by the dashed vertical line in the left panel of Fig.~\ref{im:rtanalysis_con}. As can also be seen in the right panel of Fig.~\ref{im:rtanalysis_con}, the analytical solution is indeed perfectly matched by the numerical one up to that value of $t$, where the perturbation has reached an amplitude of $\eta/\lambda\approx 0.1$. Since by then the average vertical perturbation speed, $\varv$, has not grown higher than about $9.6\times 10^{-4}\,a_{0,\textrm{max}}$, the simulated fluid should still behave as incompressible to a good approximation.

After leaving the linear regime, the perturbation growth enters a transition phase that lasts about $\unit[1.8]{s}$, until it finally reaches the non-linear regime. We define the point in time at which this happens, $t_0$, by when the perturbation amplitude has reached the value of $\lambda/2$ within a relative error of 10 per cent. We take $t_0$ (marked by the dash-dotted vertical line in Fig.~\ref{im:rtanalysis_con}) as the lower limit of the interval used for a least-squares fit of the numerical data to the analytical solution valid for the non-linear regime (Eq.~\ref{eq:fncon}). The thus obtained estimate for the rate coefficient $\alpha$ is then used in turn for plotting the analytical solution as a blue line in Fig.~\ref{im:rtanalysis_con}. The estimate, $\alpha\approx 0.0453$, has an error on the order of $10^{-4}$ per cent (like for the other two models) and lies well within the range of values given by other authors (see Sec.~\ref{sec:nonlinreg}).

\begin{figure}%[!htb]
\centering
\includegraphics[width=\columnwidth]{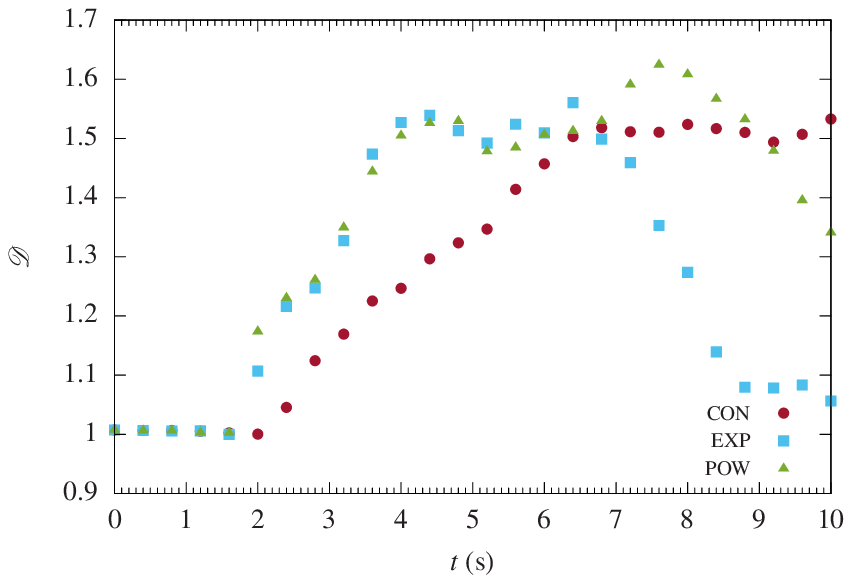}
\caption{Temporal evolution of the fractal dimension of the Rayleigh--Taylor-unstable interface for the model with constant (CON; red circles), exponential (EXP; blue squares), and power-law (POW; green triangles) vertical acceleration.}
\label{im:fracdim}
\end{figure}

Our data also allow to investigate the time evolution of the fractal or Hausdorff dimension of the RT-unstable interface. The fractal dimension is frequently defined as the exponent in the power law that relates the scale of a structure to the number of substructures required to create it, and, as such, is a quantitative way to describe the roughness of curves or surfaces. The rougher they are, the more their fractal dimension exceeds their topological dimension, $D$, taking a non-integer value in the interval between $D$ and $D+1$ \citep{Man:67}. Here, the fractal dimension is estimated using the so-called box-counting method \citep[see, e.g.][]{Pil:18}, for which we proceed as follows. The interface is overlaid by a series of ever finer regular square grids. The `boxes' that make up the coarsest of these grids have a size of $\lambda/5$, whereas the box size of the finest grid is equal to that of the cells on the finest AMR grid level. For each grid, we count the number of boxes, $N(\ell)$ ($\ell$ denotes the actual box size), within which some portion of the interface is present. Then the fractal dimension of the interface, $\mathscr{D}$, corresponds to the slope of the plot $\log N(\ell)$ vs $\log (1/\ell)$. Doing this for snapshots taken at several times, one obtains time profiles for $\mathscr{D}$, such as those shown in Fig.~\ref{im:fracdim}, with model CON being represented by the red circles. It is seen that the value of the fractal dimension remains at $\sim 1.0$ over the entire duration of the linear regime. As soon as this is left at $t\sim \unit[2]{s}$, $\mathscr{D}$ increases roughly linearly, at a rate of around $\unit[0.11]{s^{-1}}$, and then, at $t\sim\unit[6.8]{s}$ (i.e.~long after the RT instability has entered the non-linear regime at $t=t_0\sim \unit[3.8]{s}$), approaches saturation in the range of about 1.5 to 1.6. This behaviour is roughly consistent with previous analyses \citep[e.g.][]{Dub:98}.

\begin{figure}%[!htb]
\centering
\includegraphics[width=\columnwidth]{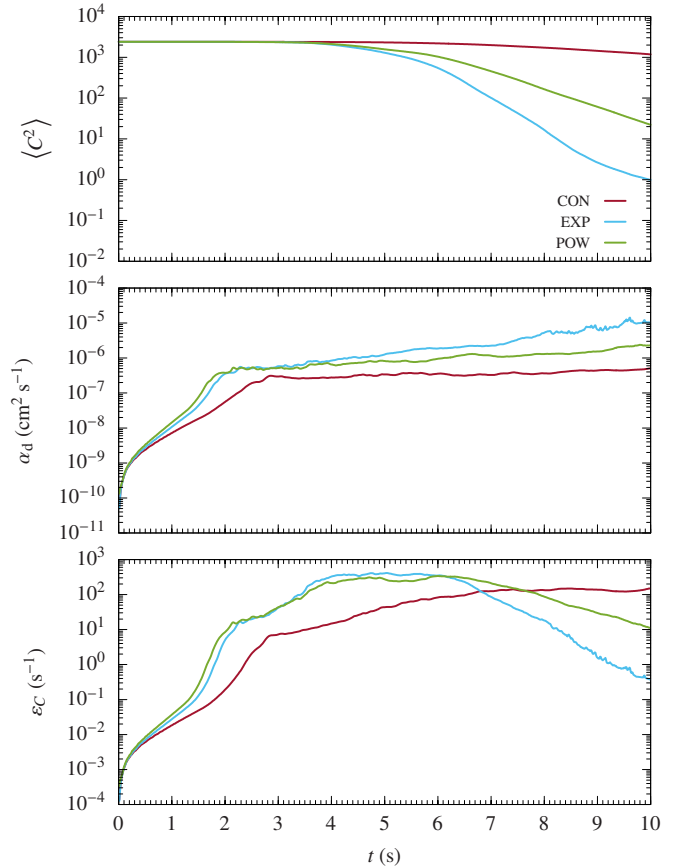}
\caption{Temporal evolution of the passive scalar variance (\emph{top panel}), diffusivity (\emph{middle panel}), and diffusion rate (\emph{bottom panel}) for the model with constant (CON; red line), exponential (EXP; blue line), and power-law (POW; green line) vertical acceleration.}
\label{im:variance}
\end{figure}

Another interesting aspect to explore is the mixing ability of the RT instability, which we do by using a passive scalar. This is a quantity that has no dynamical impact on the flow and as such obeys an advection-diffusion equation of the form \citep[see e.g.][]{Dav:15}
\begin{equation}
\frac{\partial C}{\partial t}+(\mathbfit{u}\cdot\nabla)C=\alpha_\textrm{d}\,\nabla^2 C\,,
\label{eq:advdifC}
\end{equation}
where $C$ is the scalar contaminant (e.g.~elemental concentration) and $\alpha_\textrm{d}$ is its diffusivity (which, in our case, is of purely numerical origin). Since for turbulent flows the diffusive term is usually negligible in comparison to the convective term at the scale of the large vortices, the scalar $C$ then acts like a (in our case dimensionless) marker that tags the two fluids. As initial condition we take $C_1=-49$ for the bottom fluid and $C_2=49$ for the top one, so that we start off from zero volumetric mean, $\left<C\right>=0$, which greatly simplifies the analysis. A convenient measure of the non-uniformity of $C$ is its variance, $\left<C^2\right>$, for which a dynamical equation can be derived by first multiplying both sides of Eq.~(\ref{eq:advdifC}) by $C$, leading to
\begin{equation}
\frac{\partial}{\partial t}\left(\frac{1}{2}\,C^2\right)+\nabla\cdot\left[\left(\frac{1}{2}\,C^2\right)\,\mathbfit{u}\right]=\nabla\cdot (\alpha_\textrm{d}\,C\,\nabla C)-\alpha_\textrm{d}\,(\nabla C)^2\,,
\label{eq:advdifC2}
\end{equation}
where we have taken advantage of the incompressibility condition, $\nabla\cdot\mathbfit{u}=0$. If we now assume that the distribution of $C$ is statistically homogeneous and isotropic, with the statistical properties of the flow being independent of position, which is certainly not the case during the earlier phases of the instability but becomes an increasingly better assumption the more the flow approaches the state of fully developed turbulence, taking the ensemble average of Eq.~(\ref{eq:advdifC2}) is equivalent to taking the volume average, and eliminates those terms that contain divergences. What remains is
\begin{equation}
\frac{\textrm{d}}{\textrm{d}t}\left<\frac{1}{2}\,C^2\right>=-\alpha_\textrm{d}\,\left<(\nabla C)^2\right>\,,
\end{equation}
from which follows that fluctuations in $C$ are destroyed solely by diffusion at a rate proportional to
\begin{equation}
\varepsilon_C=\alpha_\textrm{d}\,\left<(\nabla C)^2\right>\,.
\label{eq:varepsilonC}
\end{equation}
Thus when extracting the rate of change of $\left<C^2/2\right>$, and $\left<(\nabla C)^2\right>$ separately from the simulations one can obtain an estimate of $\alpha_\textrm{d}$ simply by dividing the two quantities and flipping the sign of the result. $\varepsilon_C$ then follows directly from Eq.~(\ref{eq:varepsilonC}). This procedure is illustrated by Fig.~\ref{im:variance}, which shows $\left<C^2\right>$, $\alpha_\textrm{d}$, and $\varepsilon_C$  as a function of time. The red lines correspond to model CON. As it is seen in the top panel, constant acceleration of the chosen magnitude is not sufficient in driving an instability that is able to drastically reduce the variance of a contaminant within the time frame given. Nevertheless the diffusivity of $C$ converges to a constant value of approximately $\unit[3.6\times 10^{-7}]{cm^2\,s^{-1}}$ already after about $\unit[3]{s}$ (middle panel), which lies within the transition phase of the RT instability (cf.~Fig.~\ref{im:rtanalysis_con}). Also at that time the diffusion rate $\varepsilon_C$ switches from a steep to a more gentle growth (bottom panel).  

\begin{figure}%[!htb]
\centering
\includegraphics[width=\columnwidth]{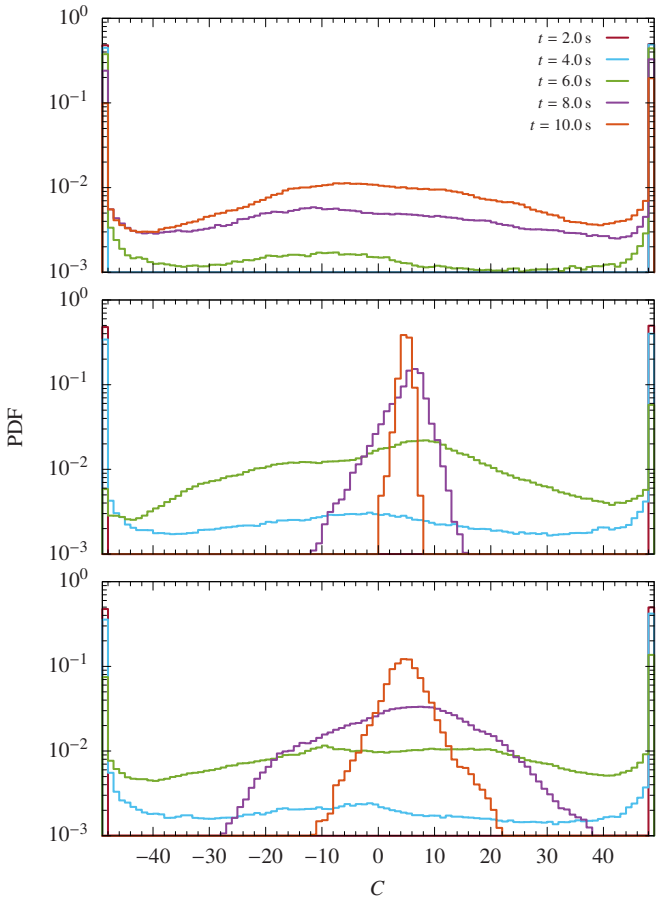}
\caption{Temporal evolution of the passive scalar probability density function (PDF) for the model with constant (CON; \emph{top panel}), exponential (EXP; \emph{middle panel}), and power-law (POW; \emph{bottom panel}) vertical acceleration.}
\label{im:pdfs}
\end{figure}

An alternative way to look at the RT mixing lies in the computation of probability density functions (PDFs) of the passive scalar field. Such PDFs are shown for several times in Fig.~\ref{im:pdfs}, with the results for model CON given in the top panel. Note that since the chosen bin width is unity, the values on the ordinate directly correspond to probabilities. It is seen that although the wings of the histograms show at all times a clear imprint of the initial data, their steady decrease nourishes the formation of a hump, whose maximum lies at intermediate values of $C$. The hump's flatness and enormous width is an indication of the still low degree of mixing of the two fluids.

%%%%%%%%%%%%%%%%%%%%%%%%%%%%%%%%%%%%%%%%%%%%%%%%%%
\subsection{Model EXP}
\label{sec:modexp}
\begin{figure*}
\resizebox{\hsize}{!}{\includegraphics{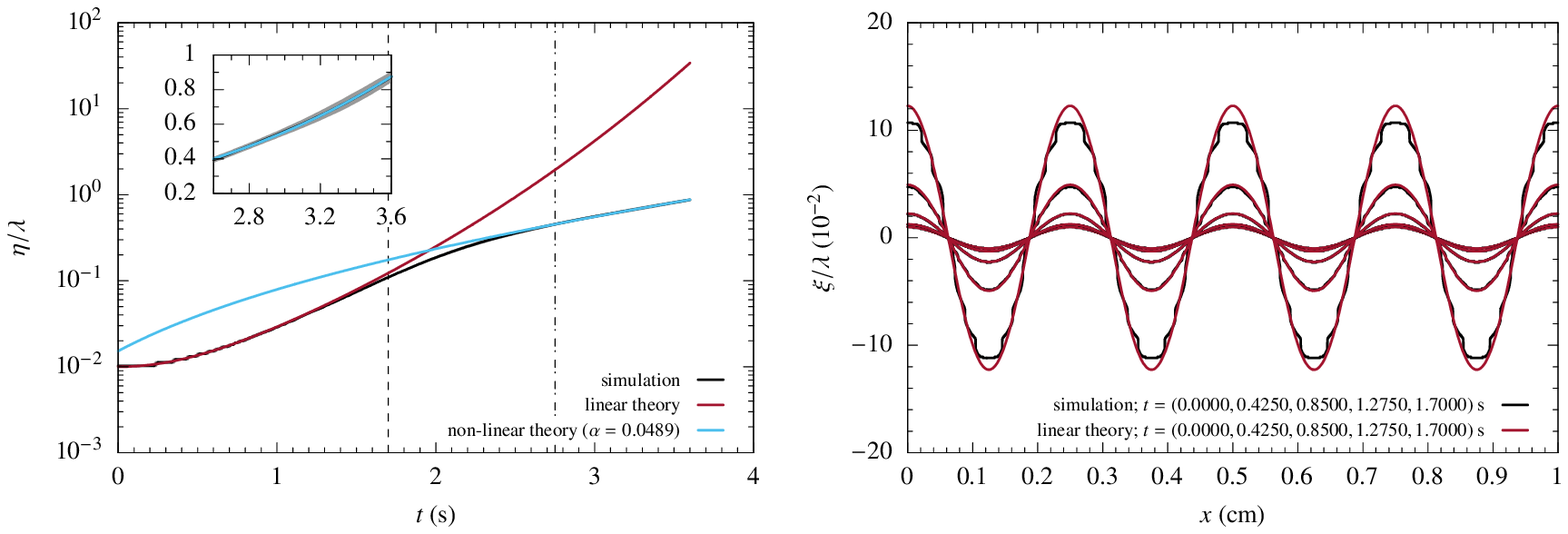}}
\caption{As for Fig.~\ref{im:rtanalysis_con}, but for model EXP.}
\label{im:rtanalysis_exp}
\end{figure*}
In the model with exponential acceleration, the RT instability develops much faster. As can be seen in Fig.~\ref{im:rtanalysis_exp}, the growth of the perturbations already reaches the end of the linear phase after about $\unit[1.7]{s}$ (at which $\varv\approx 1.2\times 10^{-3}\,a_{0,\textrm{max}}$) and enters the non-linear regime at $t\sim \unit[2.8]{s}$, with the rate coefficient taking the value $\alpha\approx 0.0489$. Accordingly, the duration of the transition phase is shortened by around 40 per cent to $\sim$$\unit[1.1]{s}$. The subsequent evolution can be nicely observed in the snapshots in the middle row of Fig.~\ref{im:timeseries}. Between $t=\unit[4]{s}$ and $\unit[6]{s}$, more precisely around $t\sim \unit[4.4]{s}$, the RT spikes reach the end of the vertical domain, as a result of which the flow field soon attains a fully turbulent state, with smaller structures and steep gradients getting smoothed out completely already before $t=\unit[8]{s}$. At $t=\unit[10]{s}$, the two fluids are almost perfectly mixed. 

Also the rate at which the fractal dimension increases from 1.0 gets significantly higher ($\sim$$\unit[0.20]{s^{-1}}$), with saturation reached at $t\sim \unit[4.4]{s}$ (blue squares in Fig.~\ref{im:fracdim}). The decrease of the profile after $t\sim\unit[6.4]{s}$ is purely artificial and non-physical, stemming from the adaptive grid used in the numerical simulations. With steep gradients getting successively removed by the turbulent flow, the grid generally coarsens. As a result, the numerical diffusion increases with a simultaneous decrease in the diffusion rate (see blue curves in the middle and bottom panel of Fig.~\ref{im:variance}) and structures on the small scales are efficiently destroyed, which is why the edge detection algorithm employed for the box-counting method eventually fails (note that if the simulation had run long enough, this drop would also have been observed in model CON). Regardless of the effects of resolution, it is due to the intermittency of the energy dissipation process, which is a direct result of vortices being teased out into finer and finer filaments by the turbulent flow, that structures cannot become completely space-filling and the fractal dimension would not increase asymptotically to 2 (or to 3, if the simulations were performed in 3D), not even for fully developed turbulence.

Quantitative evidence of the increased mixing efficiency is given by the top and middle panels of Figs.~\ref{im:variance} and \ref{im:pdfs}, respectively. The variance of the passive scalar drops below unity during the time range considered and the PDFs pile up progressively close to $C=0$. The reason why the maximum is not exactly at zero (i.e.~at the volumetric mean of the initial concentrations), but shifted to slightly higher concentrations, is probably the small but yet existing asymmetry in the growth of the RT spikes and bubbles. After a hypothetical evolution time of infinity, perfect mixing would have been achieved so that $\left<C^2\right>$ would have become zero exactly and the PDF would have degenerated into a delta function.

%%%%%%%%%%%%%%%%%%%%%%%%%%%%%%%%%%%%%%%%%%%%%%%%%%   
\subsection{Model POW}
\label{sec:modpow}
\begin{figure*}
\resizebox{\hsize}{!}{\includegraphics{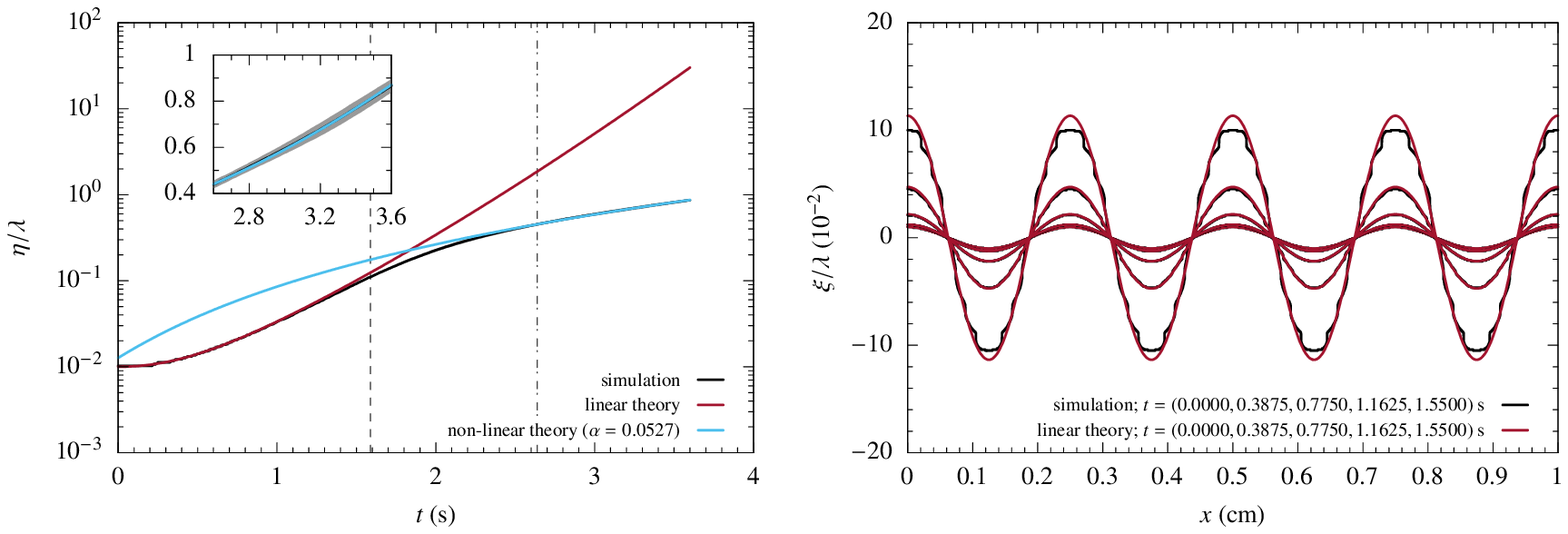}}
\caption{As for Fig.~\ref{im:rtanalysis_con}, but for model POW.}
\label{im:rtanalysis_pow}
\end{figure*}
A useful information for interpreting the results of model POW with respect to those of model EXP is that at time $t=t^*\approx\unit[2.03]{s}$ the accelerations of both models are identical, namely $g(t^*)\approx\unit[2.76]{cm\,s^{-2}}$. For $t < t^*$ the acceleration of model POW is higher than that of model EXP, and vice versa for $t > t^*$. This explains why the growth of the perturbations passes the limits of both the linear and the non-linear regime earlier than in model EXP, namely at $t\sim\unit[1.6]{s}$ (at which $\varv\approx 1.2\times 10^{-3}\,a_{0,\textrm{max}}$) and $\unit[2.6]{s}$, respectively (see Fig.~\ref{im:rtanalysis_pow}). The duration of the transition phase remains unchanged compared to model EXP ($\sim$$\unit[1.1]{s}$), whereas the rate coefficient is slightly higher ($\alpha\approx 0.0527$). The degree of turbulence attained is not sufficient, though, to diffuse away all teased out larger structures in time (see bottom row panels of Fig.~\ref{im:timeseries}), with reflection off the boundaries only happening slightly later ($t\sim\unit[4.6]{s}$) than in model EXP.

The slope of the fractal dimension increase ($\sim$$\unit[0.18]{s^{-1}}$) is comparable to that of model EXP, whereas the artificial decrease is shifted to later times, due to the ever smaller acceleration and thus mixing efficiency for $t>t^*$ (green triangles in Fig.~\ref{im:fracdim}). This is also reflected by Fig.~\ref{im:variance}, with the diffusivity values of model POW (green lines) lying at later times below that of model EXP, and the variance of $C$ in model POW never becoming as low as in model EXP. It is therefore only plausible that the PDFs of model POW (bottom panel of Fig.~\ref{im:pdfs}) never reach the narrowness of that of model EXP. 

The reason why $\alpha_\textrm{d}$ in the models POW and EXP is always higher at later times than in model CON is due to the generally coarser structure of the adaptive grid at these times.

%%%%%%%%%%%%%%%%%%%%%%%%%%%%%%%%%%%%%%%%%%%%%%%%%%
\section{Application to interstellar shells and supershells}
\label{sec:isbubbles}

%%%%%%%%%%%%%%%%%%%%%%%%%%%%%%%%%%%%%%%%%%%%%%%%%%
\subsection{Interstellar shells}
\label{sec:sphbubbles}
The hydrodynamic structure and evolution of a stellar wind bubble or an SNR that expands highly supersonically into a homogeneous or spherically symmetric medium can be quite accurately described by similarity solutions of the form
\begin{equation}
R(t)=A\,t^\beta\,.
\label{eq:simisol}
\end{equation}
Here, $A$ and the `expansion parameter' $\beta$ are positive constants, whereas $R(t)$ denotes a characteristic radius. This is usually taken to be the radius of the spheroidal (forward) shock wave that is set up by the extremely energetic mechanism at the object's centre (stellar mass loss or explosion) and that moves into the surroundings, sweeping up gas into a shell. The collision of the supersonic radial outflow with this shell leads to the formation of a second spheroidal shock wave directed into the object's interior, which is either stopped somewhere inside the object if the supersonic mass flow is continuous (`termination shock'), or it proceeds its journey to the centre if the object was created by a single explosion (`reverse shock'). In either case, the medium is reheated upon shock passage. Within the shell, a contact discontinuity separates the swept-up ISM from the ejected stellar material. 

It immediately follows from Eq.~(\ref{eq:simisol}) that the velocity, acceleration, and jerk of the forward shock, or, to a good approximation, also of the shell as a whole, are given by
\begin{align}
\dot{R}&=\beta\,A\,t^{\beta-1}=\beta\,R/t\,,\label{eq:bubblevel}\\ 
\ddot{R}&=\beta\,(\beta-1)\,A\,t^{\beta-2}=\beta\,(\beta-1)\,R/t^2\,,\\
\dddot{R}&=\beta\,(\beta-1)\,(\beta-2)\,A\,t^{\beta-3}=\beta\,(\beta-1)\,(\beta-2)\,R/t^3\,,
\end{align}
where $g\equiv\ddot{R} > 0$ (shell accelerates) for $\beta > 1$ and $\dot{g}\equiv\dddot{R} > 0$ for $\beta > 2$. Building upon the case distinction developed in Sec.~\ref{sec:linreg}, the following ranges for the density difference across the contact discontinuity and the power-law exponent constitute an RT-unstable situation:
\begin{enumerate}
\item \tabto{1cm}$(\rho_2-\rho_1)\ge -\frac{B\,(\beta-2)^2}{4\,k\,\beta\,(\beta-1)\,A\,t^\beta}$ and $\beta > 2$,
\item \tabto{1cm}$(\rho_2-\rho_1)>0$ and $1 < \beta \le 2$,
\item \tabto{1cm}cannot be fulfilled,
\item \tabto{1cm}$(\rho_2-\rho_1)<0$ and $\beta < 1$,
\end{enumerate}
with all configurations requiring that $t > 0$. Recall that the index 2 refers to the `upper' fluid layer, which, in the spheroidal context, is taken to be the one farther away from the object's centre. So in the cases (i) and (ii) the RT spikes grow inwards, whereas they grow outwards in the case (iv).

%%%%%%%%%%%%%%%%%%%%%%%%%%%%%%%%%%%%%%%%%%%%%%%%%%
\subsubsection{Circumstellar shells}
\label{sec:swbubbles}
To first approximation, the structure of the bubbles blown by the winds of massive stars (O and B types) during most of their lifetime resembles that of a volume of hot ($T_1\gtrsim\unit[10^6]{K}$) shocked wind gas enclosed by a narrow and cool ($T_2\approx\unit[10^4]{K}$) decelerating ($\beta < 1$) shell that is still exposed to the stellar ultraviolet (UV) radiation \citep{Cas:75}. The inner boundary of this thin circumstellar shell can taken to be the contact discontinuity, whereas the outer boundary is represented by the forward shock wave \citep[see][]{Dys:21}. 

Thus, since $(\rho_2-\rho_1) > 0$ across the contact discontinuity (see below), such a shell is in principle always stable against the RT instability. This would however no longer be the case if it were possible to accelerate the shell, which could be achieved either by a temporarily increasing stellar mechanical energy release rate (`mechanical luminosity') or if the system is expanding into a gas whose density is decreasing with radius, for example if the bubble breaks out of the molecular cloud in which the star was born and abruptly enters the lower-density intercloud medium \citep{Wea:77}. Following \citeauthor{Wea:77}, we assume that the wind luminosity obeys the law $L_{\star}(t)=K_{\star}\,t^m$ and the ambient density profile is given by $\rho(r)=K_\rho\,r^{-n}$. Then it is easy to show via dimensional analysis that the bubble's expansion is governed by
\begin{equation}
R(t)= C\,K_{\star}^{1/(5-n)}\,K_\rho^{-1/(5-n)}\,t^{(3+m)/(5-n)}\,,
\label{eq:wblaw2}
\end{equation}
where $C$ is a dimensionless constant and both the wind-blowing star and the ISM ahead of the shell are assumed to be at rest. As a consequence, $\dot{R}$ is the shell velocity both relative to the surrounding ISM and in the stellar frame of reference. The `usual' RT-stable state is recovered with $m=n=0$, for which $K_{\star}=L_{\star,0}$ and $K_\rho =\rho_0$, leading to
\begin{equation}
R(t)=C_0\,L_{\star,0}^{1/5}\,\rho_0^{-1/5}\,t^{3/5}\,,
\label{eq:wblaw}
\end{equation}
with $C_0=[125/(154\,\pi)]^{1/5}\approx 0.76$, as can be found through simple conservation law arguments \citep{Cas:75}. The typical bubble that surrounds an early-type star with a mass loss rate of $\dot{M}_{\star}\approx\unit[10^{-6}]{M_{\sun}\,yr^{-1}}$ and a wind velocity of $\varv_{\star}\approx\unit[2000]{km\,s^{-1}}$ is driven by a mechanical luminosity of $L_{\star,0}=\frac{1}{2}\,\dot{M}_{\star}\,\varv_{\star}^2\approx\unit[10^{36}]{erg\,s^{-1}}$. Hence at an age of $t=\unit[1]{Myr}\equiv t_0$ such a bubble has attained a radius and shell velocity of $R(t_0)\approx \unit[28]{pc}\equiv R_0$ and $\dot{R}(t_0)=3/5\,(R_0/t_0)\approx\unit[16]{km\,s^{-1}}$, respectively, when assuming the ambient ISM to have a density of $n_0 =\unit[1]{cm^{-3}}$ with solar abundances (mean molecular weight of $\mu\approx 1.30$). 

Since the time-scale for radiative cooling of a wind-blown bubble is usually shorter than its dynamical time-scale, we can take the forward shock to be isothermal. Then the density in the shell, $n_2$, is related to $n_0$ by the jump condition
\begin{equation}
n_2=\frac{\dot{R}^2}{a_2^2}\,n_0\,,
\label{eq:isothermalshock}
\end{equation}
where $a_2=\sqrt{k_\textrm{B}\,T_2/\bar{m}_2}\approx \unit[10]{km\,s^{-1}}$ is the isothermal sound speed in the shell, with a mean post-shock particle mass, $\bar{m}_2$, of about $0.62\,m_\textrm{H}$, for shells that have been completely ionized by the stellar UV photons ($k_\textrm{B}\approx \unit[1.381\times 10^{-16}]{erg\,K^{-1}}$ denotes the Boltzmann constant and $m_\textrm{H}\approx\unit[1.674\times 10^{-24}]{g}$ the hydrogen atom mass). At $t=t_0$ we hence have $n_2\approx\unit[4.4]{cm^{-3}}$. Considering that the density of the shocked wind gas is only $n_1\approx\unit[0.01]{cm^{-3}}$ \citep{Cas:75}, it holds that $\rho_1/\rho_2\ll 1$, and we can safely make the approximation $(\rho_2-\rho_1)=\rho_2\,(1-\rho_1/\rho_2)\approx\rho_2 > 0$, which should be valid even for $t\gg t_0$.

What remains to be determined is the parameter $B$. In the case of a wind-blown bubble we can set $h_1 = R$ and $h_2 = d$, which is the shell thickness. It can be found from mass conservation, noting that the mass in the thin shell must equal the mass of the ISM originally contained within the radius $R$, if the ejecta mass is negligible. Hence
\begin{equation}
4\,\pi\,R^2\,n_2\,m_\textrm{H}\,d =\frac{4\,\pi}{3}\,R^3\,n_0\,m_\textrm{H}\,,
\end{equation}
or, equivalently,
\begin{equation}
d=\frac{1}{3}\,\left(\frac{n_0}{n_2}\right)\,R\,,
\label{eq:shellthickness1}
\end{equation}
which can be combined with Eqs.~(\ref{eq:isothermalshock}), (\ref{eq:bubblevel}), and (\ref{eq:simisol}) to yield
\begin{equation}
d=\frac{a_2^2}{3\,\dot{R}^2}\,R\,=\frac{a_2^2}{3\,\beta^2\,A}\,t^{2-\beta}\,.
\label{eq:shellthickness2}
\end{equation}
Hence the shell grows in thickness with time only if $\beta <2$. The maximum size of the perturbing wavelength should be a sizable fraction of the shell radius, but not too large, that is $\lambda = f\,R$, with $f<1$. We thus have
\begin{equation}
B=\rho_1\,\coth\left(\frac{2\,\pi}{f}\,\frac{h_1}{R}\right)+\rho_2\,\coth\left(\frac{2\,\pi}{f}\,\frac{h_2}{R}\right)\,,
\label{eq:bshell}
\end{equation}
and therefore
\begin{equation}
B=\rho_2\left[\coth\left(\frac{2\,\pi}{f}\,\frac{d}{R}\right)+\frac{\rho_1}{\rho_2}\,\coth\left(\frac{2\,\pi}{f}\right)\right]\,,
\end{equation}
which, to a good approximation, is equal to $\rho_2$ for all times requiring that $f\lesssim 0.1$, as $d/R$ cannot exceed the adiabatic limit of $1/12$, posed by the corresponding compression ratio of $n_2/n_0=4$. This permits us to set $(\rho_2-\rho_1)/B=1$.

So for the RT instability to occur, $m$ and $n$ from Eq.~(\ref{eq:wblaw2}) have to satisfy either criterion (i) or (ii). As the quotient in (i) is always negative, the necessary condition for instability is $1<\beta = (3+m)/(5-n) \le 2$, and therefore either 
\begin{description}
\item $0\le m\le 2$ and $2-m < n \le (7-m)/2$, or
\item $2 < m < 7$ and $0 \le n \le (7-m)/2$, or
\item $m=7$ and $n = 0$.
\end{description}
Hence the stability criterion provided by \cite{Wea:77}, $n+m<2$, is actually incomplete.

We are now in a position to estimate the time it would take the RT instability to fragment the circumstellar shell. Replacing the constants of proportionality $K_{\star}$ and $K_\rho$ in Eq.~(\ref{eq:wblaw2}) by suitable initial data we obtain as an evolution law for the shell radius
\begin{equation}
R(t)=A_\textrm{CSS}\,t^{(3+m)/(5-n)}
\label{eq:wbevo}
\end{equation}
with
\begin{equation}
A_\textrm{CSS}=\left(\frac{125}{154\,\pi}\right)^{1/(5-n)}\,\left(\frac{L_{\star,0}}{t_0^m}\right)^{1/(5-n)}\,(\rho_0\,R_0^n)^{-1/(5-n)}\,,
\end{equation}
and hence for the shell thickness
\begin{equation}
d(t)=\frac{(5-n)^2\,a_2^2}{3\,(3+m)^2\,A_\textrm{CSS}}\,t^{(7-2\,n-m)/(5-n)}\,.
\end{equation}
For most of the time the growth of the instability should be in the non-linear regime and thus governed by Eq.~(\ref{eq:fnpoweasy}), which becomes for the above evolution law and assumptions
\begin{equation}
\begin{split}
\eta(t)&=\frac{4\,\alpha\,(m+n-2)}{3+m}\,A_\textrm{CSS}\\
&\quad\times\left[t^{(3+m)/(10-2\,n)}-t_0^{(3+m)/(10-2\,n)}\right]^2\\
&\quad+4\,\sqrt{\frac{\alpha\,(m+n-2)}{3+m}\,A_\textrm{CSS}\,\eta_0}\\
&\quad\times\left[t^{(3+m)/(10-2\,n)}-t_0^{(3+m)/(10-2\,n)}\right]+\eta_0\,,
\end{split}
\label{eq:pertcss}
\end{equation}
whereby $\eta(t_0)\equiv\eta_0=0.01\,d(t_0)$ is chosen as the initial condition. The shell should break up as soon as the perturbation amplitude has grown to the size of shell thickness, that is when $\eta(t)=d(t)$.

\begin{figure}
\centering
\includegraphics[width=\columnwidth]{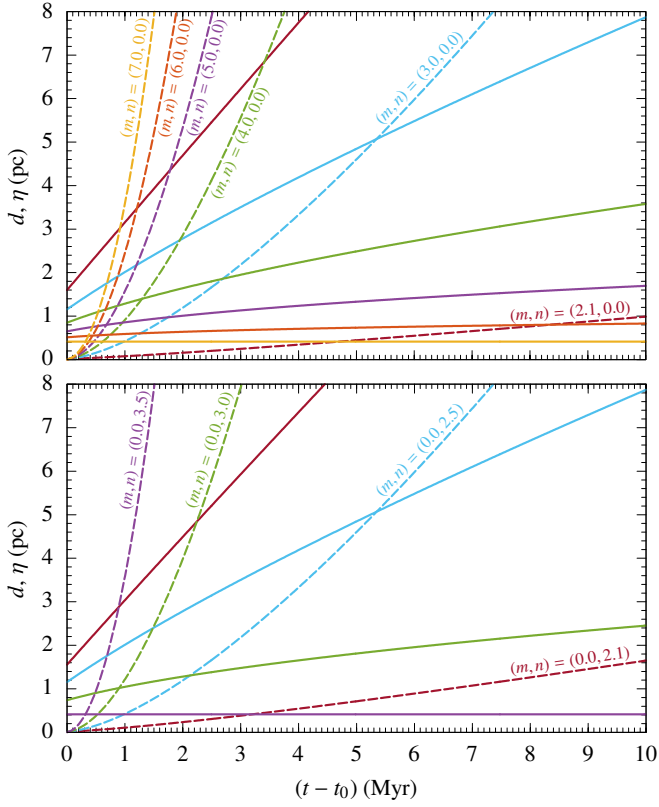}
\caption{Fragmentation time study for circumstellar shells. Shown are the shell thicknesses (solid lines) and the non-linear perturbation amplitudes (dashed lines; rate coefficient $\alpha=0.06$) as a function of time for several values of the exponents $m$ and $n$ of the power-law profiles for the stellar mechanical luminosity ($L_{\star}\propto t^m)$ and the ambient density ($\rho\propto \,r^{-n}$), respectively. In the \emph{upper (lower) panel}, the effect of the former (latter) is studied in isolation, by setting $n=0$ ($m=0$) and varying $m$ ($n$) within the Rayleigh--Taylor-unstable range. Intersections of different curves of the same color indicate shell break-ups. The time is rescaled to begin at zero.}
\label{im:windbubble}
\end{figure}

Corresponding times for shell fragmentation, as computed for several values of the exponents $m$ and $n$, and the rate coefficient $\alpha$, are given in Table~\ref{tab:breakupt_swb}. For illustration, Fig.~\ref{im:windbubble} shows the functions plotted for the case of $\alpha=0.06$. It is clearly seen that the shell integrity is much more susceptible to changes of the background density gradient (lower panel) than to variations of the wind luminosity (upper panel).

\begin{table*}
\caption{Fragmentation times for circumstellar shells, as calculated for several values, $m$ and $n$, of the power-law stellar mechanical luminosity, $L_{\star}(t)=K_{\star}\,t^m$, and ambient density profile, $\rho(r)=K_\rho\,r^{-n}$, and of the rate coefficient $\alpha$.}
\label{tab:breakupt_swb}
\begin{tabular}{ccccccc}
\hline
$m$    & $n$   & \multicolumn{5}{c}{Fragmentation time}\\
       &       & \multicolumn{5}{c}{(Myr)}\\
                 \cline{3-7}
       &       & $\alpha=0.02$ & $\alpha=0.04$ & $\alpha=0.06$ & $\alpha=0.08$ & $\alpha=0.1$\\
\hline
$2.1$  & $0.0$ & --            & --            & --            & --            & --\\
$3.0$  & $0.0$ & $29.04$       & $9.24$        & $5.35$        & $3.80$        & $2.99$\\
$4.0$  & $0.0$ & $2.99$        & $1.74$        & $1.29$        & $1.06$        & $0.91$\\
$5.0$  & $0.0$ & $1.31$        & $0.86$        & $0.68$        & $0.58$        & $0.51$\\
$6.0$  & $0.0$ & $0.79$        & $0.55$        & $0.44$        & $0.38$        & $0.34$\\
$7.0$  & $0.0$ & $0.55$        & $0.39$        & $0.32$        & $0.27$        & $0.25$\\
\hline
$0.0$  & $2.1$ & --            & --            & --            & --            & --\\
$0.0$  & $2.5$ & $29.04$       & $9.24$        & $5.35$        & $3.80$        & $2.99$\\
$0.0$  & $3.0$ & $1.86$        & $1.17$        & $0.90$        & $0.75$        & $0.66$\\
$0.0$  & $3.5$ & $0.55$        & $0.39$        & $0.32$        & $0.27$        & $0.25$\\
\hline
\end{tabular}
\end{table*}

At first glance, it may seem problematic to use relations, such as Eq.~(\ref{eq:pertcss}), that are derived from the assumption of incompressibility in an environment whose dynamics are determined by the presence of strong shock waves. This is however fully justified since we are dealing here with post-shock media whose sound speeds are several orders of magnitude higher than the speeds of the disturbances, even under the most extreme boundary conditions.

%%%%%%%%%%%%%%%%%%%%%%%%%%%%%%%%%%%%%%%%%%%%%%%%%%
\subsubsection{Shells of young supernova remnants}
\label{sec:youngsnrs}
As demonstrated by \cite{Che:82}, the earliest, so-called ejecta-dominated phase of an SNR can be described in a self-similar fashion if the density profiles of the homologously expanding (outer) ejecta and the stationary ambient medium are approximated by power laws of the form $\rho_\textrm{ej}(r,t)=K_\textrm{ej}\,r^{-n}\,t^{n-3}$ and $\rho_\textrm{a}(r)=K_\textrm{a}\,r^{-s}$, respectively, with $K_\textrm{ej}$ and $K_\textrm{a}$ being constants. The result is a shell consisting of two layers of different widths that is bounded by the forward and the reverse shock. The motion of the contact discontinuity separating the two layers is given by
\begin{equation}
R(t)=A_\textrm{SNR}\,t^{(n-3)/(n-s)}\,,
\label{eq:ysnrevolaw}
\end{equation}
with
\begin{equation}
A_\textrm{SNR}=\left(\chi\,\frac{K_\textrm{ej}}{K_\textrm{a}}\right)^{1/(n-s)}\,,
\label{eq:aysnr}
\end{equation}
where $\chi$ is a dimensionless constant that depends only on $n$ and $s$. The requirement of finite energy and mass in the flow demands that $n>5$ and $0\le s<3$. Within that range, all self-similar solutions show a steep inwardly-directed density gradient across the contact discontinuity, which, together with $\beta=(n-3)/(n-s)<1$, poses an RT-unstable configuration due to criterion (iv). 

The value of $n$ depends on the properties of the supernova (SN) progenitor and the explosion mechanism \citep{Fra:10}. As found from numerical simulations, $n=7$ describes reasonably well the ejecta structure of Type~Ia SNe, which are believed to result from the thermonuclear explosion of a CO white dwarf exceeding the Chandrasekhar limit of about $\unit[1.4]{M_{\sun}}$ through mass accretion from a companion star in a close binary system. In contrast, $n=9$--12 is considered to be a suitable approximation of the density structure of Type~II SNe, which originate from the core-collapse of a massive star \citep[see][and references therein]{Vin:12}. 

For a homogeneous ambient medium $s=0$ and $K_\textrm{a}=\rho_0$, whereas for a SN exploding inside a steady-state stellar wind, mass conservation implies that $s=2$ and
\begin{equation}
K_\textrm{a}=\frac{\dot{M}_{\star}}{4\,\pi\,\varv_{\star}}\,.
\label{eq:kaysnr}
\end{equation}
These cases should be characteristic for Type~Ia and Type~II SNe, respectively. Among the most extensively studied of each type are the young SNRs Tycho and Cas~A. Canonical values for their age, explosion energy, $E_\textrm{SN}$, and ejecta mass, $M_\textrm{ej}$, are given in Table~\ref{tab:inputysnr}.

\begin{table*}
\caption{Model parameters for young supernova remnants (SNRs) of Type~Ia (Tycho-like) and Type~II (Cas~A-like), with (approximate) ages, explosion energies ($E_\textrm{SN}$), and ejecta masses ($M_\textrm{ej}$) taken from the literature. $n$ and $s$ are the exponents in the power-law density profiles for the ejecta ($\rho_\textrm{ej}\propto r^{-n}$) and the ambient medium ($\rho_\textrm{a}\propto r^{-s}$), respectively. The parameter $\chi$ results from integrating the fluid equations as outlined in {\protect\cite{Che:82}}. The same holds for the width of the inner and outer shell layer, $h_1$ and $h_2$, normalized to the radius $R$ of the contact discontinuity between the two media, with $K_d$ being the sum of the ratios. The parameter $A_\textrm{SNR}$ that completes the self-similar evolution law (\ref{eq:ysnrevolaw}) is calculated from Eq.~(\ref{eq:aysnr}).}
\label{tab:inputysnr}
\begin{tabular}{ccccccccccc}
\hline
SNR type & Age    & $E_\textrm{SN}$         & $M_\textrm{ej}$ & $n$ & $s$ & $\chi$  & $h_1/R$ & $h_2/R$ & $K_d$   & $A_\textrm{SNR}$\\
         & (yr)   & ($\unit[10^{51}]{erg}$) & (M$_{\sun}$)    &     &     &         &         &         &         & (cgs units)\\
\hline
Tycho    & $450$  & $1.0$                   & $1.4$           & $7$ & $0$ & $1.198$ & $0.065$ & $0.181$ & $0.246$ & $1.5\times 10^{13}$\\
Cas~A    & $340$  & $2.2$                   & $2.0$           & $9$ & $2$ & $0.096$ & $0.019$ & $0.250$ & $0.269$ & $1.8\times 10^{10}$\\
\hline
\end{tabular}
\end{table*}

Now for studying the RT instability that operates on the shell of a Tycho-like SNR we follow the approach of \cite{Che:82} and assume that the density profile $\rho\propto r^{-7}$ only holds for the outer $3/7$ of the progenitor star (by mass). This allows us to calculate the constant $K_\textrm{ej}$ through
\begin{equation}
K_\textrm{ej}=\frac{25\,E_\textrm{SN}^2}{21\,\pi\,M_\textrm{ej}}\,.
\end{equation}
As suggested by observations \citep{Tia:11}, the ambient ISM is taken to be homogeneous with density $\rho_0=K_\textrm{a}=\unit[0.6\,m_\textrm{H}]{cm^{-3}}$ \citep[cf.][]{Wil:20}. The value of the remaining parameter necessary for calculating $A_\textrm{SNR}$ via Eq.~(\ref{eq:aysnr}), $\chi$, is obtained by numerically integrating the governing fluid equations written in suitable similarity variables separately for both the inner and the outer shell layer, with boundary conditions imposed by the reverse shock, the contact discontinuity, and the forward shock. Table~\ref{tab:inputysnr} gives the results of this procedure. With the ratios $h_1/R$ and $h_2/R$ also obtained in this way, the parameter $B$ can be estimated using Eq.~(\ref{eq:bshell}). Since $\rho_2/\rho_1\ll 1$ in the shells of young SNRs, we find that $B\approx \rho_1$, and thus $(\rho_2-\rho_1)/B\approx -1$, when allowing for perturbation wavelengths of $0.1\,R$ or smaller. Besides considering the growth of the instability relative to that of the individual shell layer widths $h_1$ and $h_2$ (with $h_2/h_1\approx 2.8$ for Tycho and $13.2$ for Cas~A), we are also interested in the evolution of the shell's full thickness,
\begin{equation}
d(t)=K_d\,A_\textrm{SNR}\,t^{(n-3)/(n-s)}\,,
\end{equation}
which poses an upper limit for the time until a possible shell break-up; $K_d$ is as given in Table~\ref{tab:inputysnr}, noting that $d=h_1+h_2$. Perturbations should grow as (replacing $K$ in Eq.~\ref{eq:fnpoweasy} by $\beta\,(\beta-1)\,A_\textrm{SNR}$)
\begin{equation}
\begin{split}
\eta(t)&=\frac{4\,\alpha\,(3-s)}{n-3}\,A_\textrm{SNR}\\
&\quad\times\left[t^{(n-3)/(2\,n-2\,s)}-t_0^{(n-3)/(2\,n-2\,s)}\right]^2\\
&\quad+4\,\sqrt{\frac{\alpha\,(3-s)}{n-3}\,A_\textrm{SNR}\,\eta_0}\\
&\quad\times\left[t^{(n-3)/(2\,n-2\,s)}-t_0^{(n-3)/(2\,n-2\,s)}\right]+\eta_0\,,
\end{split}
\label{eq:pertsnr}
\end{equation}
where we again take $\eta(t_0)\equiv \eta_0=0.01\,d(t_0)$ as initial condition, with $t_0=\unit[10]{yr}$ for both SNRs.

To carry out the analogous analysis for a Cas~A-like SNR, the only difference lies in the determination of the values of $K_\textrm{ej}$ and $K_\textrm{a}$. The former can be found by supposing the ejecta to be composed of two regions: a core of uniform density and an enveloping region for which the aforementioned density power law applies \citep{Blo:01}. The core's radius at which the two density profiles merge must then expand with the constant velocity
\begin{equation}
\varv_\textrm{c}=\sqrt{\frac{(10\,n-50)\,E_\textrm{SN}}{(3\,n-9)\,M_\textrm{ej}}}\,,
\end{equation}
so that $K_\textrm{ej}$ takes the form
\begin{equation}
K_\textrm{ej}=\frac{5\,n-25}{2\,\pi\,n}\,E_\textrm{SN}\,\varv_\textrm{c}^{n-5}\,.
\end{equation}
Following \cite{Che:03} and \cite{Lam:03}, we adopt for Cas~A the value $n=9$. For the calculation of $K_\textrm{a}$ via Eq.~(\ref{eq:kaysnr}) we suppose that the medium that surrounds Cas~A has been shaped by the red supergiant wind of its progenitor star, with the wind parameter values $\dot{M}_{\star}=\unit[1.54\times 10^{-5}]{M_{\sun}\,yr^{-1}}$ and $\varv_{\star}=\unit[4.7]{km\,s^{-1}}$ as given by the stellar evolution models of \cite{Hir:04}. This assumption is again supported by observations \citep[e.g.][]{Lee:14}.

\begin{figure}%[!htb]
\centering
\includegraphics[width=\columnwidth]{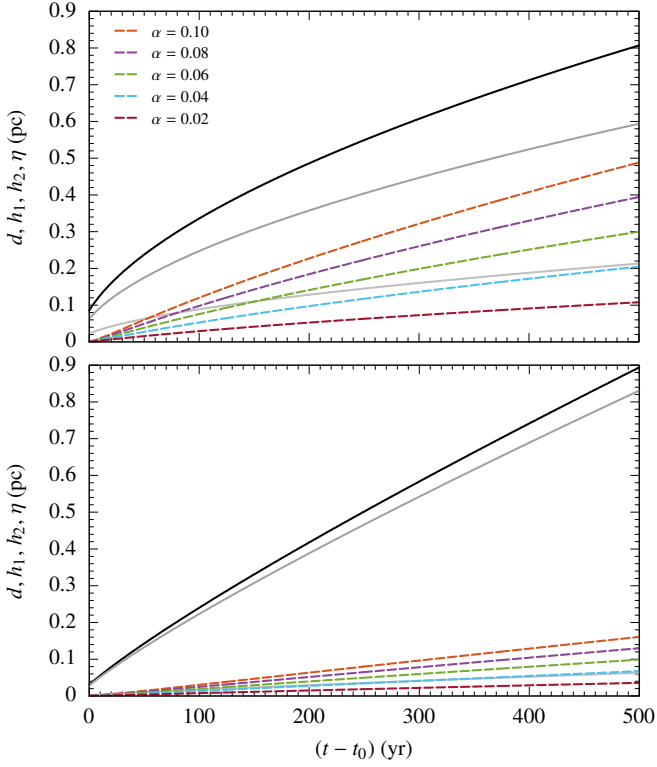}
\caption{Temporal evolution of the Rayleigh--Taylor instability within the shell of a Tycho-like (\emph{top panel}) and Cas~A-like (\emph{bottom panel}) supernova remnant. Shown are the total thickness of the shell (black solid line), the widths of its inner and outer layer (light grey and dark grey solid line, respectively), and the non-linear perturbation amplitudes for several values of the rate coefficient $\alpha$ (dashed lines) as a function of time. Note that in the bottom plot the light grey solid line is almost exactly covered by the blue dashed line. The time is rescaled to begin at zero.}
\label{im:ysnr}
\end{figure}

The resulting `race' between the shell thicknesses (solid lines) and perturbation amplitudes calculated for several values of the rate coefficient $\alpha$ (dashed lines) is depicted for both SNRs in Fig.~\ref{im:ysnr}. As can be seen, the RT spikes are unable to completely penetrate either the outer shell layer of Tycho or of Cas~A within a timespan of $\unit[500]{yr}$, which lies above the current ages of both SNRs. Nevertheless, the perturbations are generally much more elongated for Tycho. What can also be noticed is that if the growth of the RT bubbles is described by a rate coefficient of about 0.04 or higher -- which, according to experiments \citep{Dim:04,Ban:20}, might actually be the case -- at least the inner layer of the shells of both SNRs would get completely penetrated.

Although the detailed modeling of a specific SNR is beyond the scope of this paper, our results for the inner layer are consistent with X-ray observations of both Tycho \citep{War:05} and Cas~A \citep{Pat:09}. For the outer layer, though, the same observations report on a reduced width when compared to the model of \cite{Che:82}, which, as simulations indicate, might be due to efficient particle acceleration at the forward shock \citep{Fer:10,War:13}, and even more, at least locally, due to initial asymmetries from the SN itself \citep{Orl:16,Fer:19}. Besides altering the density gradient, this makes it easier for the RT spikes to come closer to the forward shock, or even deform it. However, these modifications of the classical idealized picture should primarily be reflected in the value of the parameter $K_d$, leaving the growth curves of the perturbations obtained with Eq.~(\ref{eq:pertsnr}) relatively unchanged.

It is believed that the RT instability does not accompany SNRs throughout their entire evolution. The instability should die out when the shell behind the forward shock becomes cool and dense as a result of catastrophic cooling, thus reversing the density gradient across the contact discontinuity \citep[see e.g.][]{Dys:21}.

%%%%%%%%%%%%%%%%%%%%%%%%%%%%%%%%%%%%%%%%%%%%%%%%%%
\subsection{Interstellar supershells}
\label{sec:sush}

Herinafter, axisymmetric cylindrical coordinates $(r,z)$ are employed, with $z$ being measured along the axis of symmetry. Concerning the nomenclature, we denote dimensionless variables by a tilde, where all quantities with the dimension of length are normalized to the scale or characteristic height $H$ (surfaces and volumes correspondingly to $H^2$ and $H^3$).

%%%%%%%%%%%%%%%%%%%%%%%%%%%%%%%%%%%%%%%%%%%%%%%%%%
\subsubsection{Generic Kompaneets superbubbles with a time-dependent energy input rate}
\label{sec:komp}

\defcitealias{Bau:13}{BB13}

Shells whose diameters exceed the characteristic scale of the gas distribution in the vertical direction of the galactic disc can no longer be spherical but must be elongated in this direction. The energy that fuels the dynamics of such `supershells', which are omnipresent in star-forming galaxies, could be provided by the joint action of stellar winds and coherent Type~II SN explosions in OB associations, harboring several dozens of massive stars. As long as the time between successive explosions is short in comparison to the characteristic evolution time-scale of the SB -- as it is usually the case for a few to a few tens of million years -- the energy input can be assumed to be continuous and the SB can be thought of a gigantic wind-blown bubble. However, the additional length-scale introduced by the density scale height is the reason why evolved SBs cannot be treated in a self-similar fashion.

The first model that was able to overcome this problem was introduced by \cite{Kom:60}, yet in the context of atomic-weapons research. Like the wind-bubble model by \cite{Cas:75} and \cite{Wea:77}, it is based on the strong-shock approximation, implying that the external pressure is negligible in comparison to the internal pressure, which is taken to be uniform across the entire SB volume. This assumption is motivated by the high temperatures and thus sound speeds in the SB interior that allow for redistribution of the internal energy to a nearly isobaric state before the shock front moves an appreciable distance \citep{Bis:95}. Another assumption of the Kompaneets approximation is that every element of the shock surface expands into the direction of the shock normal, which should be particularly true for the vertices of the SB. Making use of all these assumptions, the shape and the time evolution of the SB can be determined analytically.

Contrary to the original model, which was tailored to be applied to ground bursts on Earth, we place the explosion centre either in a midplane ($\tilde{z}=0$) with the density (scaled to the midplane density value $\rho_0$) decreasing symmetrically away from it in an exponential manner (model I),
\begin{equation}
\tilde{\rho}_\textrm{I}(\tilde{z})=\exp(-|\tilde{z}|)\,,
\label{eq:dprofI}
\end{equation}
or shift it at an arbitrary height $\tilde{z}_0$ above the base of an exponential atmosphere (normalized to the density $\rho_{\tilde{z}_0}$ at $\tilde{z}=\tilde{z}_0$; model II),
\begin{equation}
\tilde{\rho}_\textrm{II}(\tilde{z})=\exp(-\tilde{z}+\tilde{z}_0)\,.
\label{eq:dprofII}
\end{equation}
Also unlike the original model, not a single explosion is considered, but a time-dependent energy input rate due to sequential SN explosions of massive stars according to a galactic initial mass function (IMF). Both extensions were already presented by \citet[hereafter \citetalias{Bau:13}]{Bau:13} and aim at modelling the SB evolution under more realistic conditions. In contrast to this earlier work, however, when investigating the RT instability in the supershell, we do not limit ourselves to the short linear phase, but rather follow the instability's long-term development by considering specifically the non-linear phase, and, in particular, take into account the supershell's time-dependent acceleration. For the sake of a self-contained presentation, we briefly review the results obtained by \citetalias{Bau:13} that are relevant for the present analysis.

The shapes of the SBs that arise from applying the Kompaneets formalism on the density profiles (\ref{eq:dprofI}) and (\ref{eq:dprofII}) can be demonstrated to be
\begin{equation}
\tilde{r}_\textrm{I}(\tilde{y},\tilde{z})=
\begin{cases}
2\,\arccos\left[\frac{1}{2}\,\e^{\tilde{z}/2}\,\left(1-\frac{\tilde{y}^2}{4}+\e^{-\tilde{z}}\right)\right]& \tilde{z} \ge 0\,, \\
2\,\arccos\left[\frac{1}{2}\,\e^{-\tilde{z}/2}\,\left(1-\frac{\tilde{y}^2}{4}+\e^{\tilde{z}}\right)\right]& \tilde{z} < 0\,,
\end{cases}
\end{equation}
and
\begin{equation}
\tilde{r}_\textrm{II}(\tilde{y},\tilde{z})=2\,\arccos\left[\frac{1}{2}\,\e^{(\tilde{z}-\tilde{z}_0)/2}\,\left(1-\frac{\tilde{y}^2}{4}+\e^{-\tilde{z}+\tilde{z}_0}\right)\right]\,,
\end{equation}
respectively, where $\tilde{y}$ denotes a transformed time variable (with the dimension of a length). Setting $\tilde{r}_{\textrm{I},\textrm{II}}=0$, we obtain as expressions for the top and bottom vertex positions
\begin{align}
\tilde{z}_{\textrm{t},\textrm{I}}(\tilde{y})&=-2\,\ln(1-\tilde{y}/2)\label{eq:ztI}\,,\\
\tilde{z}_{\textrm{t},\textrm{II}}(\tilde{y})&=-2\,\ln(1-\tilde{y}/2)+\tilde{z}_0\label{eq:ztII}\,,
\end{align}
and
\begin{align}
\tilde{z}_{\textrm{b},\textrm{I}}(\tilde{y})&=-\tilde{z}_{\textrm{t},\textrm{I}}(\tilde{y})\label{eq:zbI}\,,\\
\tilde{z}_{\textrm{b},\textrm{II}}(\tilde{y})&=-2\,\ln(1+\tilde{y}/2)+\tilde{z}_0\label{eq:zbII}\,,
\end{align}
respectively. Note that for $\tilde{y}\rightarrow 2$, the magnitudes of Eqs.~(\ref{eq:ztI})--(\ref{eq:zbI}) tend to infinity, which implies that parts of the supershell experience infinite acceleration at a finite time. This unphysical behaviour is due to the steep exponential decline of the background media applied.

To simplify the further analysis, we follow \citetalias{Bau:13} and approximate the shapes of the SBs by prolate ellipsoids. Since the SB in model I is mirror-symmetrical with respect to the midplane, each half of the SB can be covered by an ellipsoid with a semi-major axis of
\begin{equation}
\tilde{a}_\textrm{I}(\tilde{y})=\frac{\tilde{z}_{\textrm{t},\textrm{I}}(\tilde{y})-\tilde{z}_{\textrm{b},\textrm{I}}^{*}(\tilde{y})}{2}
=2\,\arctanh(\tilde{y}/2)\,,
\end{equation}
where $\tilde{z}_{\textrm{b},\textrm{I}}^{*}(\tilde{y})=-2\,\ln(1+\tilde{y}/2)$ corresponds to the bottom of an unshifted SB in a pure exponential atmosphere. For model II, a single ellipsoid with
\begin{equation}
\tilde{a}_\textrm{II}(\tilde{y})=\frac{\tilde{z}_{\textrm{t},\textrm{II}}(\tilde{y})-\tilde{z}_{\textrm{b},\textrm{II}}(\tilde{y})}{2}
=2\,\arctanh(\tilde{y}/2)
\end{equation}
is sufficient for the entire SB. The centres of the ellipsoids located on the positive half of the $z$-axis are at
\begin{equation}
\tilde{z}_{\textrm{c},\textrm{I},\textrm{II}}(\tilde{y})=\tilde{z}_{\textrm{t},\textrm{I},\textrm{II}}(\tilde{y})-\tilde{a}_{\textrm{I},\textrm{II}}(\tilde{y})\,.
\end{equation}
Thus
\begin{equation}
\tilde{z}_{\textrm{c},\textrm{I}}(\tilde{y})=-\ln(1-\tilde{y}^2/4)
\end{equation}
and
\begin{equation}
\tilde{z}_{\textrm{c},\textrm{II}}(\tilde{y})=\tilde{z}_0-\ln(1-\tilde{y}^2/4)\,.
\end{equation}
The semi-minor axes should be set equal to the maximum half-width extensions of the SBs in $r$-direction (derivable via $\partial \tilde{r}_{\textrm{I},\textrm{II}}/\partial \tilde{z}=0$), which are the same for both models,
\begin{equation}
\tilde{b}_{\textrm{I},\textrm{II}}(\tilde{y})=2\,\arcsin(\tilde{y}/2)\,.
\end{equation}
Now everything is known to calculate the evolution of the ellipsoids through
\begin{equation}
\tilde{r}_{\textrm{ell},\textrm{I},\textrm{II}}(\tilde{y},\tilde{z})=\sqrt{\tilde{b}_{\textrm{I},\textrm{II}}^2(\tilde{y})-\frac{\tilde{b}_{\textrm{I},\textrm{II}}^2(\tilde{y})}{\tilde{a}^2_{\textrm{I},\textrm{II}}(\tilde{y})}\,[\tilde{z}-\tilde{z}_{\textrm{c},\textrm{I},\textrm{II}}(\tilde{y})]^2}\,.
\label{eq:evoell}
\end{equation}
The SB volumes at a given `time' $\tilde{y}$ can then be written as
\begin{equation}
\begin{split}
\tilde{V}_{\textrm{I}}(\tilde{y})&=2\,\pi\,\int_{0}^{\tilde{z}_{\textrm{t},\textrm{I}}(\tilde{y})}\tilde{r}_{\textrm{ell},\textrm{I}}^2(\tilde{y},\tilde{z})\,\textrm{d}\tilde{z}\\
&=2\,\pi\,\left\{\tilde{b}_\textrm{I}^2(\tilde{y})\,\tilde{z}_{\textrm{t},\textrm{I}}(\tilde{y}) -\frac{\tilde{b}_\textrm{I}^2(\tilde{y})}{\tilde{a}_\textrm{I}^2(\tilde{y})}\right.\\
&\quad\times\left.\left[\frac{\tilde{z}_{\textrm{t},\textrm{I}}^3(\tilde{y})}{3}-\tilde{z}_{\textrm{c},\textrm{I}}(\tilde{y})\,\tilde{z}_{\textrm{t},\textrm{I}}^2(\tilde{y})+\tilde{z}_{\textrm{c},\textrm{I}}^2(\tilde{y})\,\tilde{z}_{\textrm{t},\textrm{I}}(\tilde{y})       \right]  \right\}
\label{eq:VI}
\end{split}
\end{equation}
and
\begin{equation}
\begin{split}
\tilde{V}_{\textrm{II}}(\tilde{y})&=\frac{4\,\pi}{3}\,\tilde{a}_\textrm{II}(\tilde{y})\,\tilde{b}_\textrm{II}^2(\tilde{y})\\
&=\frac{32\,\pi}{3}\,\arcsin^2(\tilde{y}/2)\,\arctanh(\tilde{y}/2)\,.
\end{split}
\end{equation}

In a power-law prescription for the time-dependent mechanical luminosity of the form
\begin{equation}
L_\textrm{SN}(t)=K_\textrm{SN}\,t^\delta\,,
\end{equation}
the exponent $\delta=-(\Gamma/\nu+1)$ contains the slope of the IMF, $\Gamma$, and $\nu=1.628$, which is a parameter in a simple fitting law for the main-sequence lifetime of massive stars, $\tau_\textrm{MS}(M)=\kappa\,M^{-\nu}$ (stellar masses $M$ in units of solar masses; $\kappa=\unit[1.3\times 10^9]{yr}$ for $7\le M\le 85$), as derived from the stellar evolution models of \cite{Eks:12}. The rate coefficient $K_\textrm{SN}$ is given by
\begin{equation}
K_\textrm{SN}=\frac{N_0\,E_\textrm{SN}\,\kappa^{\Gamma/\nu}}{\nu}\,,
\end{equation}
where the normalisation coefficient of the IMF, $N_0$, is chosen such that there is exactly one star in the highest-mass bin,
\begin{equation}
N_0=\frac{\Gamma}{M_\textrm{u}^\Gamma-(M_\textrm{u}-1)^\Gamma}\,,
\end{equation}
with every mass bin containing an integer number of stars and $M_\textrm{u}$ denoting the upper mass limit of the IMF. The value of $M_\textrm{u}$ is either dictated by observations of the specific star cluster under consideration or has to be postulated. The number of all massive stars in the OB association is then simply given by
\begin{equation}
N_\star=\frac{N_0\,(M_\textrm{u}^\Gamma-M_\textrm{l}^\Gamma)}{\Gamma}\,,
\end{equation}
where $M_\textrm{l}=\unit[8]{M_{\sun}}$ is the lower initial mass limit for SN progenitor stars. Various values for the slope of the IMF for massive stars can be found in the literature: $\Gamma=-1.15$ \citep{Bal:03}, $-1.35$ \citep{Sal:55}, $-1.7$ \citep{Sca:86}, among others.

The transformation of $\tilde{y}$ to physical times is done through
\begin{equation}
\begin{split}
\tilde{t}(\tilde{y})&=\left[\frac{(\delta+3)^2\,(7\,\delta+11)}{20\,\bar{\gamma}^2}\right.\\
&\quad\left.\times\left(\int_0^{\tilde{y}}\sqrt{\tilde{V}_{\textrm{I},\textrm{II}}(\tilde{y}')}\,\textrm{d}\tilde{y}'\right)^2\right]^{1/(\delta+3)}\,,
\label{eq:ty}
\end{split}
\end{equation}
with $\tilde{t}(\tilde{y})$ given in units of the characteristic SB evolution time-scale
\begin{equation}
\tau_\textrm{SB}=\left(\frac{\rho_{0,\tilde{z}_0}\,H^5}{K_\textrm{SN}}\right)^{1/(\delta + 3)}\,,
\label{eq:sbevots}
\end{equation}
and $\bar{\gamma}=\sqrt{(\gamma^2-1)/2}$. Instead of evaluating the integral in Eq.~(\ref{eq:ty}) numerically, we expand it into a power series to seventh order, giving for model I
\begin{equation}
\begin{split}
\int_0^{\tilde{y}}\sqrt{\tilde{V}_\textrm{I}(\tilde{y}')}\,\textrm{d}\tilde{y}'&= 0.8187\,\tilde{y}^{5/2}+0.1096\,\tilde{y}^{7/2}\\
&\quad +0.0299\,\tilde{y}^{9/2}+0.0085\,\tilde{y}^{11/2}\\
&\quad +0.0028\,\tilde{y}^{13/2}+\mathscr{O}(\tilde{y}^{15/2})\,,
\end{split}
\end{equation}
and for model II
\begin{equation}
\begin{split}
\int_0^{\tilde{y}}\sqrt{\tilde{V}_\textrm{II}(\tilde{y}')}\,\textrm{d}\tilde{y}'&= 0.8187\,\tilde{y}^{5/2}+0.0379\,\tilde{y}^{9/2}\\
&\quad +0.0037\,\tilde{y}^{13/2}+\mathscr{O}(\tilde{y}^{15/2})\,.
\end{split}
\end{equation}

Both the velocity and the acceleration, and thus also the growth rate of the RT instability, should always be highest at the top vertex of the supershell (for model I also at the bottom vertex). The formula for the velocity there (normalized to $H\,\tau_\textrm{SB}^{-1}$) can be shown to be
\begin{equation}
\dot{\tilde{z}}_\textrm{t}(\tilde{y})=\zeta\,\tilde{X}(\tilde{y})\,\tilde{Y}(\tilde{y})\,\tilde{Z}(\tilde{y})\,,
\label{eq:sushvel}
\end{equation}
whereas we have for the acceleration (normalized to $H\,\tau_\textrm{SB}^{-2}$)
\begin{equation}
\begin{split}
\ddot{\tilde{z}}_\textrm{t}(\tilde{y})&=\zeta^2\,\tilde{X}(\tilde{y})\,\tilde{Y}(\tilde{y})\,\left[\tilde{X}'(\tilde{y})\,\tilde{Y}(\tilde{y})\,\tilde{Z}(\tilde{y})+\tilde{X}(\tilde{y})\,\tilde{Y}'(\tilde{y})\,\tilde{Z}(\tilde{y})\right.\\
&\quad\left.+\tilde{X}(\tilde{y})\,\tilde{Y}(\tilde{y})\,\tilde{Z}'(\tilde{y})\right]\,,
\label{eq:sushacc}
\end{split}
\end{equation}
with
\begin{equation}
\zeta=\bar{\gamma}\,\sqrt{\frac{5}{7\,\delta +11}}\,\left[\frac{(\delta +3)^2\,(7\,\delta+11)}{20\,\bar{\gamma}^2}\right]^{(\delta+1)/[2\,(\delta+3)]}\,,
\end{equation}
\begin{equation}
\tilde{X}(\tilde{y})=\left(\int_0^{\tilde{y}}\sqrt{\tilde{V}_{\textrm{I},\textrm{II}}(\tilde{y}')}\,\textrm{d}\tilde{y}'\right)^{(\delta+1)/(\delta+3)}\,,
\end{equation}
\begin{equation}
\tilde{Y}(\tilde{y})=\frac{1}{\sqrt{\tilde{V}_{\textrm{I},\textrm{II}}(\tilde{y})}}\,,
\end{equation}
and
\begin{equation}
\tilde{Z}(\tilde{y})=\frac{1}{1-\tilde{y}/2}\,.
\end{equation}
Assuming the supershell to be thin in comparison to the extension of the SB, and to be composed exclusively of swept-up gas, its thickness can again be estimated from mass conservation, leading to
\begin{equation}
\begin{split}
\tilde{d}_{\textrm{I}}(\tilde{y})&=\tilde{M}_{\textrm{sh},\textrm{I}}(\tilde{y})\left/\left[2\,\pi\,\displaystyle\int_{0}^{\tilde{z}_{\textrm{t},\textrm{I}}(\tilde{y})}4\,\exp(-\tilde{z})\right.\right.\\
&\quad\times\left.\tilde{r}_{\textrm{ell},\textrm{I}}(\tilde{y},\tilde{z})\,\sqrt{1+\left(\pde{\tilde{r}_{\textrm{ell},\textrm{I}}}{\tilde{z}}\right)^2}\,\textrm{d}\tilde{z}\right]
\label{eq:sushthicknessI}
\end{split}
\end{equation}
and
\begin{equation}
\begin{split}
\tilde{d}_{\textrm{II}}(\tilde{y})&=\tilde{M}_{\textrm{sh},\textrm{II}}(\tilde{y})\left/\left[2\,\pi\,\int_{\tilde{z}_{\textrm{b},\textrm{II}}(\tilde{y})}^{\tilde{z}_{\textrm{t},\textrm{II}}(\tilde{y})}4\,\exp(-\tilde{z}+\tilde{z}_0)\right.\right.\\
&\quad\times\left.\tilde{r}_{\textrm{ell},\textrm{II}}(\tilde{y},\tilde{z})\,\sqrt{1+\left(\pde{\tilde{r}_{\textrm{ell},\textrm{II}}}{\tilde{z}}\right)^2}\,\textrm{d}\tilde{z}\right]\,,
\label{eq:sushthicknessII}
\end{split}
\end{equation}
with $\tilde{M}_{\textrm{sh},\textrm{I}}(\tilde{y})=\pi\,\int_{0}^{\tilde{z}_{\textrm{t},\textrm{I}}(\tilde{y})} \exp(-\tilde{z})\,\tilde{r}_{\textrm{ell},\textrm{I}}^2(\tilde{y},\tilde{z})\,\textrm{d}\tilde{z}$ and $\tilde{M}_{\textrm{sh},\textrm{II}}(\tilde{y})=\pi\,\int_{\tilde{z}_{\textrm{b},\textrm{II}}(\tilde{y})}^{\tilde{z}_{\textrm{t},\textrm{II}}(\tilde{y})} \exp(-\tilde{z}+\tilde{z}_0)\,\tilde{r}_{\textrm{ell},\textrm{II}}^2(\tilde{y},\tilde{z})\,\textrm{d}\tilde{z}$ denoting the supershell's half and total mass, respectively. The factor 4 in the denominators of Eqs.~(\ref{eq:sushthicknessI}) and (\ref{eq:sushthicknessII}) originates from the compression ratio of a strong adiabatic shock.

It is due to the density stratification of disc galaxies that the tip of a supershell only begins to accelerate at a certain point in time, $t_0$ (before that it decelerates). The value of $\tilde{y}=\tilde{y}_0$ that corresponds to this time is given by the zero of Eq.~(\ref{eq:sushacc}). The value of $N_0$, and thus of $M_\textrm{u}$ and $N_{\star}$, is chosen to be consistent with a supershell expanding already at $t_0$ at a velocity equal to three times the isothermal sound speed, $a=\sqrt{k_\textrm{B}\,T/\bar{m}}\approx\unit[6]{km\,s^{-1}}$, with $T=\unit[6000]{K}$ and $\bar{m}=1.3\,m_\textrm{H}$. In this way, the strong-shock assumption on which the Kompaneets approximation is based is ensured a posteriori. An overview of the parameters that result for the background models I and II, and the various IMF slopes, is given in Table \ref{tab:accsush}. The small deviations from the data presented in \citetalias{Bau:13} are solely due to the update made on the stellar lifetime model in the present paper.

\begin{table*}
\caption{Characteristic values of accelerating supershells. 
Rows 1--3: symmetric background (BG) model consisting of a Lockman layer \citep{Loc:84} with scale height $H=\unit[500]{pc}$ and Galactic midplane density $n_0=\unit[0.5]{cm^{-3}}$; rows 4--6: off-plane model with the star cluster located at $z_0=0.7\,H$ above the base of a low-scale height ($H=\unit[100]{pc}$), high-density ($n_0=\unit[10]{cm^{-3}}$) pure exponential atmosphere \citepalias[cf.][]{Bau:13}. 
$\Gamma$:~slope of the initial mass function (IMF); 
$\tilde{y}_0$:~transformed time at which the supershell’s top vertex begins to accelerate; 
$\tilde{z}_\textrm{t}(\tilde{y}_0)$:~top vertex position of the supershell at the `time' $\tilde{y}_0$; $
M_\textrm{u}$:~upper mass limit of the IMF; 
$N_{\star}$:~minimum number of stars in the mass range $[\unit[8]{M_{\sun}},M_\textrm{u}]$ required for the supershell to accelerate; 
$\tau_\textrm{SB}$:~characteristic superbubble (SB) evolution time-scale; 
$\tilde{t}(\tilde{y}_0)\equiv \tilde{t}_0$:~age of the SB when its shell begins to accelerate; 
$\tilde{d}(\tilde{y}_0)$:~supershell thickness at the age $\tilde{t}_0$.}
\label{tab:accsush}
\begin{tabular}{ccccccccc}
\hline
BG model & $\Gamma$ & $\tilde{y}_0$ & $\tilde{z}_\textrm{t}(\tilde{y}_0)$ & $M_\textrm{u}$ & $N_{\star}$ & $\tau_\textrm{SB}$ & $\tilde{t}(\tilde{y}_0)$ & $\tilde{d}(\tilde{y}_0)$ \\
         &          &               &                                     & (M$_{\sun}$)   &             & (Myr)              &                          &                          \\
\hline
I        & $-1.15$  & $1.090$       & $1.58$                              & $28$           & $77$        & $16.46$            & $1.88$                   & $0.12$\\
I        & $-1.35$  & $1.037$       & $1.46$                              & $27$           & $76$        & $16.56$            & $1.77$                   & $0.11$\\
I        & $-1.70$  & $0.943$       & $1.28$                              & $24$           & $76$        & $16.77$            & $1.58$                   & $0.10$\\
\hline
II       & $-1.15$  & $1.018$       & $2.12$                              & $13$           & $7$         & $3.78$             & $1.50$                   & $0.08$\\
II       & $-1.35$  & $0.963$       & $2.01$                              & $14$           & $10$        & $3.76$             & $1.42$                   & $0.08$\\
II       & $-1.70$  & $0.865$       & $1.83$                              & $16$           & $19$        &  $3.75$            & $1.27$                   & $0.07$\\
\hline
\end{tabular}
\end{table*}

As for circumstellar shells it should hold to good approximation that $(\rho_2-\rho_1)/B = 1$, which means that the temporal evolution of the RT instability during the non-linear regime is obtained from (numerically) solving the ordinary differential equation
\begin{equation}
\dot{\tilde{\eta}}(\tilde{t})=2\,\sqrt{\alpha\,\ddot{\tilde{z}}_\textrm{t}(\tilde{y}(\tilde{t}))\,\tilde{\eta}(\tilde{t})}\,,
\label{eq:fnsush}
\end{equation}
where $\tilde{y}(\tilde{t})$ is the inverse function of $\tilde{t}(\tilde{y})$, and $\tilde{\eta}(\tilde{t}_0)=0.01\,d(\tilde{y}(\tilde{t}_0))$ is set to be the initial condition.

\begin{figure}%[!htb]
\centering
\includegraphics[width=\columnwidth]{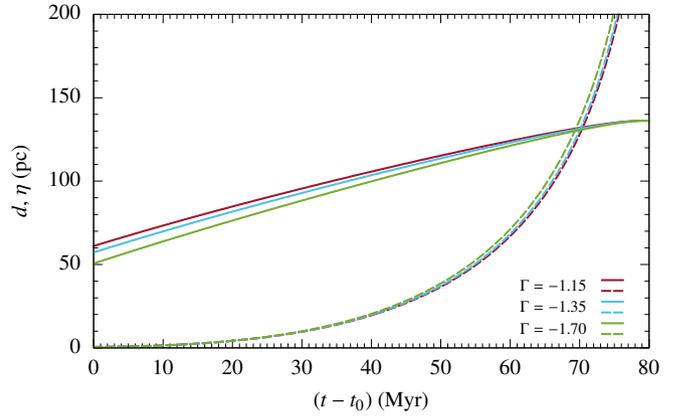}
\caption{Fragmentation time study for supershells exposed to background model I. Shown are the shell thicknesses (solid line) and the non-linear perturbation amplitudes (dashed lines; rate coefficient $\alpha=0.06$) as a function of time for several values, $\Gamma$, of the slope of the initial mass function. Intersections of different curves of the same color indicate supershell break-ups due to Rayleigh--Taylor instabilities. The time is rescaled to begin at zero.}
\label{im:sushI}
\end{figure}
\begin{figure}%[!htb]
\centering
\includegraphics[width=\columnwidth]{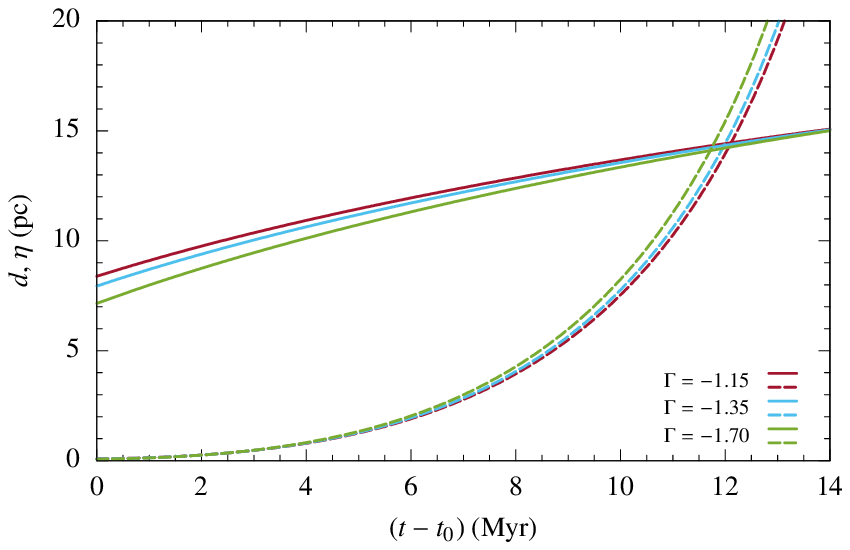}
\caption{As for Fig.~\ref{im:sushI}, but for background model II.}
\label{im:sushII}
\end{figure}
\begin{table*}
\caption{Fragmentation times and superbubble ages at shell break-up, as calculated for several values of the rate coefficient $\alpha$, and slopes, $\Gamma$, of the initial mass function. The background (BG) models are the same as in Table~\ref{tab:accsush}.}
\label{tab:breakupt_sush}
\begin{tabular}{ccccccccccccc}
\hline
BG model & $\Gamma$ & \multicolumn{5}{c}{Fragmentation time}                                       && \multicolumn{5}{c}{Age at break-up}\\
         &          & \multicolumn{5}{c}{(Myr)}                                                    && \multicolumn{5}{c}{(Myr)}\\
           \cline{3-7}                                                                                \cline{9-13}
         &          & $\alpha=0.02$ & $\alpha=0.04$ & $\alpha=0.06$ & $\alpha=0.08$ & $\alpha=0.1$ && $\alpha=0.02$ & $\alpha=0.04$ & $\alpha=0.06$ & $\alpha=0.08$ & $\alpha=0.1$\\
\hline
I        & $-1.15$  & $81.13$       & $75.54$       & $70.44$       & $66.16$       & $62.70$      && $112.13$      & $106.54$      & $101.44$      & $97.16$       & $93.70$\\
I        & $-1.35$  & $80.98$       & $75.35$       & $70.05$       & $65.74$       & $62.10$      && $110.35$      & $104.72$      & $99.42$       & $95.12$       & $91.47$\\
I        & $-1.70$  & $80.68$       & $74.65$       & $69.28$       & $64.75$       & $61.06$      && $107.20$      & $101.16$      & $95.79$       & $91.26$       & $87.57$\\
\hline
II       & $-1.15$  & $15.23$       & $13.37$       & $12.13$       & $11.18$       & $10.46$      && $20.89$       & $19.04$       & $17.79$       & $16.85$       & $16.13$\\
II       & $-1.35$  & $15.13$       & $13.24$       & $11.97$       & $11.02$       & $10.31$      && $20.46$       & $18.58$       & $17.30$       & $16.36$       & $15.64$\\
II       & $-1.70$  & $14.92$       & $13.01$       & $11.70$       & $10.76$       & $10.05$      && $19.68$       & $17.76$       & $16.45$       & $15.51$       & $14.80$\\
\hline
\end{tabular}
\end{table*}

Solutions to Eq.~(\ref{eq:fnsush}) calculated for both background models, and the various IMF slopes, are plotted together with the time profiles for the supershell thickness in Figs.~\ref{im:sushI} and \ref{im:sushII}, with the rate coefficient $\alpha$ fixed to $0.06$. The fragmentation times, which correspond to the intersections of the profiles, are listed in Table \ref{tab:breakupt_sush}, together with those obtained for other values of $\alpha$ in the range $0.02$--$0.1$. The ages of the SBs at the time of shell break-up are shown as well. As expected, the growth of the RT instability can be significantly boosted by a high value of $\alpha$. By contrast, the profiles are remarkably insensitive to the IMF chosen, with steeper IMF slopes decreasing the fragmentation time by only a few per cent, at most. And, when looking at the influence of the background model, it is obvious that off-plane explosions (model II) dramatically increase the growth rate of the instability (in those parts of the supershell that are farthest away from the midplane) -- we are talking about a factor of 5--6 for the model parameters selected!

Leaving aside minor differences in the stellar lifetime prescription, it is important to conclude that current models, which do not explicitly take into account the time-dependence of the instability growth rate, underestimate the fragmentation time of the supershell by up to a factor of a few, even for the highest values of the efficiency parameter $\alpha$ \citepalias[see e.g.][]{Bau:13}.

%%%%%%%%%%%%%%%%%%%%%%%%%%%%%%%%%%%%%%%%%%%%%%%%%%
\subsubsection{The eROSITA bubbles}
\label{sec:erosita}
Recently, as a prominent example, \citet{Pre:20} have reported the existence of the EBs, observed in an X-ray survey by the \emph{eROSITA} satellite, extending about $\unit[14]{kpc}$ perpendicular to both sides of the Galactic disc and emanating from the GC region. In fig.~3 of their paper it can be seen that the conspicuous EBs in the energy range 0.3--$\unit[1]{keV}$ neatly enclose the FBs observed in gamma rays in the GeV energy range, suggesting that relativistic electrons boost low-energy photons. If the bubbles are the result of explosive events in the GC region, such as SNe, stellar tidal disruption events (TDEs) or active galactic nucleus (AGN) jet activity, these electrons could quite naturally be accelerated by shock waves involved. The energy input rate inferred for the EBs of $\sim$$\unit[10^{41}]{erg\,s^{-1}}$ is consistent with that estimated for the FBs (which is a factor of 3--10 lower than for the EBs) by \citet{Ko:20}.

Since the Galactic Centre Black Hole (GCBH) is rather quiescent now, it seems likely that the energy input is episodic, mimicking a stellar wind or SB. In such a scenario one would expect a termination shock, which cannot be seen in the X-ray data. Therefore the activity must have been high in the past, but the last episode of energy input must have occurred a while ago, so that the termination shock must have degraded into a sound wave. This could explain the relative thickness of the X-ray emitting shells when compared to the tenuous hot interior outlined by the FBs. The relative smoothness of the shells argues for the time-dependent RT instability not having set in yet.

Due to their considerable size, the gas distribution in the halo should have been decisive for the evolution of the EBs for a long time. Unfortunately, like the Galactic disc, the gas distribution for the halo is hard to constrain. For instance, \cite{Cor:91} and \cite{Bis:18} suggested an exponential profile as in model I (Eq.~\ref{eq:dprofI}), whereas \cite{Mil:13,Mil:16} proposed a so-called $\beta$-model profile, which takes the form
\begin{equation}
\tilde{\rho}_\textrm{III}(\tilde{z})=(1+|\tilde{z}|)^{-3\,\beta}
\label{eq:dprofIII}
\end{equation}
for the GC region (hereinafter referred to as model III). If $\beta=2/3$, one finds that within the Kompaneets framework the shock front would evolve as
\begin{equation}
\tilde{r}_\textrm{III}(\tilde{y},\tilde{z})=
\begin{cases}
\sqrt{\sinh^2\tilde{y}-(1+\tilde{z}-\cosh\tilde{y})^2}& \tilde{z} \ge 0\,, \\
\sqrt{\sinh^2\tilde{y}-(1-\tilde{z}-\cosh\tilde{y})^2}& \tilde{z} < 0\,.
\end{cases}
\end{equation}
Assuming a constant energy input rate in the range $L_\textrm{GC}=3\times (10^{40}$--$\unit[10^{41})]{erg\,s^{-1}}$ requires setting $\delta=0$ and $K_{\textrm{SN}}=L_\textrm{GC}$ in Eqs.~(\ref{eq:ty}) and (\ref{eq:sbevots}).

If, as in Sec.~\ref{sec:komp}, the shape of the SB is approximated by prolate ellipsoids, one can describe its evolution in the two aforementioned halo profiles by Eq.~(\ref{eq:evoell}) and (\ref{eq:VI}), respectively -- in the former just replace the subscript `I,II' by `I,III', and in the latter `I' by `I,III'. The expressions to be used for model III are
\begin{equation}
\tilde{a}_\textrm{III}(\tilde{y})=\tilde{b}_\textrm{III}(\tilde{y})=\sinh\tilde{y}
\end{equation}
and
\begin{equation}
\tilde{z}_{\textrm{c},\textrm{III}}(\tilde{y})=\cosh\tilde{y}-1\,,
\end{equation}
since
\begin{equation}
\tilde{z}_{\textrm{t},\textrm{III}}(\tilde{y})=\cosh\tilde{y}+\sinh\tilde{y}-1
\end{equation}
and
\begin{equation}
\tilde{z}_{\textrm{b},\textrm{III}}^{*}(\tilde{y})=\cosh\tilde{y}-\sinh\tilde{y}-1\,.
\end{equation}

Hence, contrary to model I, the SB parts above and below the midplane maintain a spherical shape, actually very similar to the shape observed for the EBs, with the top of the SB (or any other part) never reaching infinity in finite time. In order to account for this fact, and to keep the error low, the integral  $\int_0^{\tilde{y}}\sqrt{\tilde{V}_{\textrm{I},\textrm{III}}(\tilde{y}')}\,\textrm{d}\tilde{y}'$ that occurs in several formulae is expanded to the seventh order for model I (as in Sec.~\ref{sec:komp}) and to the 14th order for model III. In the case of the latter, the first few terms read
\begin{equation}
\begin{split}
\int_0^{\tilde{y}}\sqrt{\tilde{V}_\textrm{III}(\tilde{y}')}\,\textrm{d}\tilde{y}'&= 0.8187\,\tilde{y}^{5/2}+0.2193\,\tilde{y}^{7/2}\\
&\quad +0.0817\,\tilde{y}^{9/2}+0.0214\,\tilde{y}^{11/2}\\
&\quad +0.0052\,\tilde{y}^{13/2}+\mathscr{O}(\tilde{y}^{15/2})\,.
\end{split}
\end{equation}

The relations for the velocity (Eq.~\ref{eq:sushvel}) and acceleration (Eq.~\ref{eq:sushacc}) of the top vertex of the supershell are also valid for model III, with the only change that
\begin{equation}
\tilde{Z}(\tilde{y})=\sinh\tilde{y}+\cosh\tilde{y}\,.
\end{equation}

The formula describing the thickness of the supershell for model III is then
\begin{equation}
\begin{split}
\tilde{d}_{\textrm{III}}(\tilde{y})&=\tilde{M}_{\textrm{sh},\textrm{III}}(\tilde{y})\left/\left\{2\,\pi\,\int_{0}^{\tilde{z}_{\textrm{t},\textrm{III}}(\tilde{y})}4\,(1+\tilde{z})^{-2}\right.\right.\\
&\quad\times\left.\tilde{r}_{\textrm{ell},\textrm{III}}(\tilde{y},\tilde{z})\,\sqrt{1+\left(\pde{\tilde{r}_{\textrm{ell},\textrm{III}}}{\tilde{z}}\right)^2}\,\textrm{d}\tilde{z}\right\}\,,
\label{eq:sushthicknessIII}
\end{split}
\end{equation}
with $\tilde{M}_{\textrm{sh},\textrm{III}}(\tilde{y})=\pi\,\int_{0}^{\tilde{z}_{\textrm{t},\textrm{III}}(\tilde{y})}(1+\tilde{z})^{-2}\,\tilde{r}_{\textrm{ell},\textrm{III}}^2(\tilde{y},\tilde{z})\,\textrm{d}\tilde{z}$ denoting the shell's half-mass. Taking the forward shock to be adiabatic should be justified when considering the very long cooling time (on the order of $\unit[10^{8}]{yr}$) for the X-ray emitting gas observed downstream of the shock. The rather low upstream Mach number of $\mathscr{M}\approx 1.5$, as estimated from the Rankine-Hugoniot condition for the temperature increase from about $\unit[0.2]{keV}$ outside of the EBs to around $\unit[0.3]{keV}$ inside \citep{Pre:20}, may be the absolute lower limit of what can still be treated with the Kompaneets approximation, which actually presumes a strong shock. Both background models fulfill this requirement, with the accelerating supershell remaining always faster than three times the isothermal sound speed in the halo of $a\approx\unit[115]{km\,s^{-1}}$ (assuming that $T=\unit[10^6]{K}$ and $\bar{m}=0.62\,m_\textrm{H}$). Further characteristic values are listed in Table \ref{tab:accerosita}.

\begin{figure}%[!htb]
\centering
\includegraphics[width=\columnwidth]{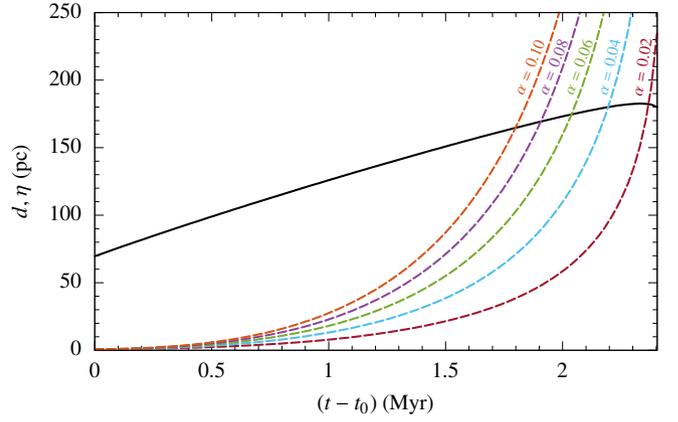}
\caption{Fragmentation time study for the \textit{eROSITA} bubbles exposed to background model I and a constant Galactic Centre (GC) mechanical luminosity of $L_\textrm{GC}=\unit[3\times 10^{41}]{erg\,s^{-1}}$. Shown are the shell thickness (solid line) and the non-linear perturbation amplitudes for several values of the rate coefficient $\alpha$ (dashed lines) as a function of time. Intersections of the dashed lines with the solid line indicate shell break-ups due to Rayleigh--Taylor instabilities. The time is rescaled to begin at zero.}
\label{im:erositaIbu}
\end{figure}

\begin{figure}%[!htb]
\centering
\includegraphics[width=\columnwidth]{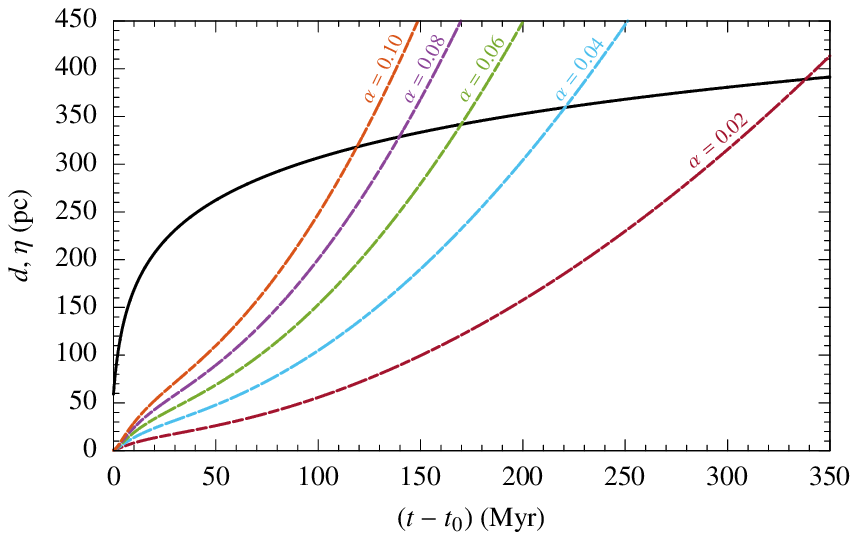}
\caption{As for Fig.~\ref{im:erositaIbu}, but for background model III.}
\label{im:erositaIIIbu}
\end{figure}

\begin{figure}%[!htb]
\centering
\includegraphics[width=\columnwidth]{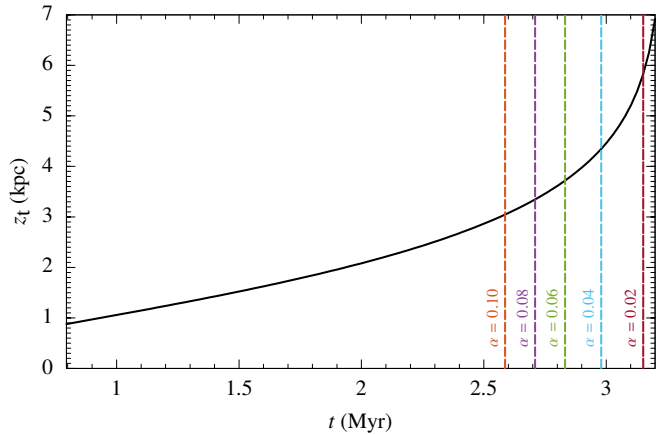}
\caption{Position of the top vertex of the \textit{eROSITA} bubbles exposed to background model I and a constant Galactic Centre (GC) mechanical luminosity of $L_\textrm{GC}=\unit[3\times 10^{41}]{erg\,s^{-1}}$ as a function of the superbubble age (solid line). The dashed lines mark the ages at which the supershells break up due to Rayleigh--Taylor instabilities, as calculated for several values of the rate coefficient $\alpha$.}
\label{im:erositaIsize}
\end{figure}

\begin{figure}%[!htb]
\centering
\includegraphics[width=\columnwidth]{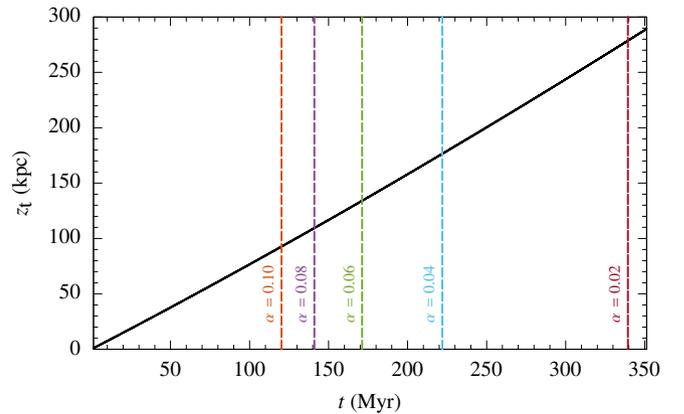}
\caption{As for Fig.~\ref{im:erositaIsize}, but for background model III.}
\label{im:erositaIIIsize}
\end{figure}

Modelling the interaction of the RT instability with the supershell analogous to Sec.~\ref{sec:komp}, we obtain the results compiled in Table \ref{tab:breakupt_erosita} and \ref{tab:vertexpos_erosita}. Since for the exponential halo (see also Figs.~\ref{im:erositaIbu} and \ref{im:erositaIsize}) the RT instability would break up the outer wall of the EBs already before they could reach their present-day extension, which is not indicated by the data of \cite{Pre:20}, this background model can be ruled out, at least for the lower halo. By the same token, we can conclude from the power-law halo model (see also Figs.~\ref{im:erositaIIIbu} and \ref{im:erositaIIIsize}) that the EBs cannot grow older than about $\unit[730]{Myr}$, possibly rather $\sim$$\unit[480]{Myr}$, when relying on the latest (terrestrial) measurements of the rate coefficient $\alpha$ for RT bubbles \cite[see][and references therein]{Ban:20}. This is still well above the current age of the EBs, which, taking into account their current extension above and below the Galactic disc ($\sim$$\unit[14]{kpc}$), we find to be about $\unit[20]{Myr}$ (see Fig.~\ref{im:erositaIIIsize}). 

If the EBs are driven by mechanical luminosities as high as $\unit[3\times 10^{41}]{erg\,s^{-1}}$, which is theoretically achievable by TDEs or AGN-like activities associated with the GCBH, their maximum final age would be further reduced to the range 120--$\unit[220]{Myr}$. In either case, the EBs would have only just reached about 15 per cent (or even less) of their final vertical extension (see Fig.~\ref{tab:vertexpos_erosita}), provided that the power of the energy source in the GC remains unchanged in the future.
\begin{table*}
\caption{Characteristic values of the \textit{eROSITA} bubbles. Both background (BG) models applied are symmetric, consisting either of an exponential halo with scale height $H=\unit[670]{pc}$ and Galactic midplane density $n_0=\unit[0.03]{cm^{-3}}$ \citep[model I;][]{Nor:92} or of a power-law halo with characteristic height $H=\unit[260]{pc}$ and $n_0=\unit[0.46]{cm^{-3}}$ \citep[model III;][]{Mil:13}. 
$L_\textrm{GC}$:~mechanical luminosity due to the energy source situated in the Galactic Centre (GC) region; 
$\tilde{y}_0$:~transformed time at which the supershell's top vertex begins to accelerate; 
$\tilde{z}_\textrm{t}(\tilde{y}_0)$:~top vertex position of the supershell at the `time' $\tilde{y}_0$; 
$\tau_\textrm{SB}$:~characteristic superbubble (SB) evolution time-scale for $L_\textrm{GC}=\unit[3\times 10^{40}]{erg\,s^{-1}}$ ($\unit[3\times 10^{41}]{erg\,s^{-1}}$); 
$\tilde{t}(\tilde{y}_0)\equiv \tilde{t}_0$:~age of the SB when its shell begins to accelerate; 
$\tilde{d}(\tilde{y}_0)$:~supershell thickness at the age $\tilde{t}_0$.}
\label{tab:accerosita}
\begin{tabular}{ccccccc}
\hline
BG model & $L_\textrm{GC}$                      & $\tilde{y}_0$ & $\tilde{z}_\textrm{t}(\tilde{y}_0)$ & $\tau_\textrm{SB}$ & $\tilde{t}(\tilde{y}_0)$ & $\tilde{d}(\tilde{y}_0)$ \\
         & $(\unit[10^{40}]{erg\,s^{-1}}$)       &               &                                     & (Myr)              &                          &                          \\
\hline
I        & $3$ $(30)$                            & $0.962$       & $1.31$                              & $1.06$ $(0.49)$    & $1.62$                   & $0.10$\\
\hline
III      & $3$ $(30)$                            & $1.584$       & $3.87$                              & $0.54$ $(0.25)$    & $5.00$                   & $0.23$\\
\hline
\end{tabular}
\end{table*}
\begin{table*}
\caption{Fragmentation times and ages at shell break-up for the \textit{eROSITA} bubbles, as calculated for several values of the rate coefficient $\alpha$, and for high and low (constant) Galactic Centre (GC) mechanical luminosities, $L_\textrm{TDE}$. The background (BG) models are the same as in Table~\ref{tab:accerosita}.}
\label{tab:breakupt_erosita}
\begin{tabular}{ccccccccccccc}
\hline
BG model & $L_\textrm{GC}$   & \multicolumn{5}{c}{Fragmentation time}                                       && \multicolumn{5}{c}{Age at break-up}\\
         & (erg\,s$^{-1}$)    & \multicolumn{5}{c}{(Myr)}                                                    && \multicolumn{5}{c}{(Myr)}\\
                                \cline{3-7}                                                                      \cline{9-13}
         &                    & $\alpha=0.02$ & $\alpha=0.04$ & $\alpha=0.06$ & $\alpha=0.08$ & $\alpha=0.1$ && $\alpha=0.02$ & $\alpha=0.04$ & $\alpha=0.06$ & $\alpha=0.08$ & $\alpha=0.1$\\
\hline
I        & $3\times 10^{40}$  & $5.07$        & $4.70$        & $4.39$        & $4.12$        & $3.86$       && $6.79$        & $6.42$        & $6.10$        & $5.84$        & $5.57$\\
I        & $3\times 10^{41}$  & $2.36$        & $2.18$        & $2.04$        & $1.91$        & $1.79$       && $3.15$        & $2.98$        & $2.83$        & $2.71$        & $2.59$\\
\hline
III      & $3\times 10^{40}$  & $728.35$      & $475.23$      & $365.92$      & $300.85$      & $256.12$     && $731.06$      & $477.95$      & $368.63$      & $303.57$      & $258.83$\\
III      & $3\times 10^{41}$  & $338.07$      & $220.58$      & $169.85$      & $139.64$      & $118.88$     && $339.33$      & $221.84$      & $171.10$      & $140.90$      & $120.14$\\
\hline
\end{tabular}
\end{table*}
\begin{table*}
\caption{Top vertex position of the \textit{eROSITA} bubbles at the time of shell break-up, as calculated for several values of the rate coefficient $\alpha$. The background (BG) models are the same as in Table~\ref{tab:accerosita}.}
\label{tab:vertexpos_erosita}
\begin{tabular}{cccccc}
\hline
BG model & \multicolumn{5}{c}{Top vertex position at break-up}\\
         & \multicolumn{5}{c}{(kpc)}\\
         \cline{2-6}                                                                     
         & $\alpha=0.02$ & $\alpha=0.04$ & $\alpha=0.06$ & $\alpha=0.08$ & $\alpha=0.1$\\
\hline
I        & $5.82$        & $4.34$        & $3.72$        & $3.34$        & $3.05$\\
\hline
III      & $278.64$      & $176.41$      & $134.03$      & $109.35$      & $92.63$\\
\hline
\end{tabular}
\end{table*}

Even when only reaching the lowest possible age at break-up ($\sim$$\unit[120]{Myr}$ for $\alpha=0.1$), the thermal energy content of the EBs that would result for $L_\textrm{GC}=\unit[3\times 10^{40}]{erg\,s^{-1}}$ ($\sim$$\unit[2.5\times 10^{56}]{erg\,s^{-1}}$) and for $L_\textrm{GC}=\unit[3\times 10^{41}]{erg\,s^{-1}}$ ($\sim$$\unit[1.1\times 10^{57}]{erg\,s^{-1}}$) is still higher than their currently emitted X-ray energy \citep[$\sim$$1.3\times 10^{56}$\,erg;][]{Pre:20}. This implies that all lifetimes (and thus final sizes) that can be derived from background model III for this range of mechanical luminosities are principally consistent with the observations.

We emphasize that the results presented here assume a steady energy release in the GC region. If the energy input is instead episodic, which is not unlikely for TDEs, and extremely likely for AGNs due to their intrinsic variability, much shorter lifetimes and much smaller final sizes are theoretically achievable.

\section{Conclusions}
\label{sec:concl}
Sudden energy releases associated with strong time-variable pressure gradients in inhomogeneous plasmas are common in astrophysics. As we demonstrated, such jerks have noticeable effects on media with density gradients, requiring a treatment of time-dependent RT instability. We examined this problem in detail, both analytically and numerically, taking into account all possible variations of the ambient medium, from constant to power-law, and eventually to exponential atmospheres. As typical examples, we discussed stellar wind bubbles, young SNRs, and SBs. 

Our results can be summarized as follows:
\begin{enumerate}
\item the RT instability must be treated time-dependently whenever the jerk time-scale falls below the dynamical time-scale;
\item even for the ordinary RT instability the exponential growth is limited by a second inhibiting term; 
\item with an increase in acceleration the instability develops even faster into the non-linear regime, in which it remains for most of the time;
\item in the non-linear phase the RT instability is characterized by spike and bubble structures, which experience some differences in buoyancy;
\item the instability shows ergodic behaviour and can be followed numerically to Hausdorff dimensions of 1.6, only limited by the numerical scheme;
\item the velocity field shows a steady increase in vorticity, which is characteristic for the development of turbulent flows;
\item a power-law or exponential increase in acceleration leads to PDFs for a passive scalar which are peaked in contrast to the classical constant acceleration case, clearly indicating that mixing is faster in case of a RT instability driven by a jerk;
\item our analytical analysis, stretching from the linear growth to the fully non-linear regime, is nicely corroborated by our numerical simulations with only a slight gap between the end of the linear and the beginning of the non-linear regime; this small gap could be presumably bridged by an extension to fundamental mode coupling like in the analysis of \cite{Liu:20};
\item the theoretical analysis was applied to a number of important astrophysical situations, in particular stellar wind bubbles, Tycho- and Cas A-like SNRs, representative for young Type Ia and II SNe, and to more general Kompaneets solutions, appropriate for SBs breaking out of the Galactic disc;
\item the analysis of the shells of these objects clearly demonstrates an increase of the instability growth rate with increasing acceleration, leading to an earlier fragmentation of the shells; the growth rate was found to be generally higher in the shells of wind-blown bubbles embedded in a medium with a power-law radial density decline than in those with power-law rising stellar mechanical luminosities,  Tycho-like than Cas~A-like SNRs, and one-sided than symmetric SBs;
\item the RT instability operating within the shell of the EBs might restrict their lifetime to the range 120--$\unit[220]{Myr}$, which implies that they are currently only about 15 per cent (or even less) of their final size; much shorter lifetimes and much smaller final sizes might however be achieved if the energy input in the GC region occurs not in a constant but in an episodic fashion, which is not unlikely for TDEs, and extremely likely for AGNs due to their intrinsic variability; the spherical shape of the EBs argues for the $\beta$-model description of the gas distribution in the Galactic halo of \cite{Mil:13,Mil:16}, for which we estimate the current age of the EBs to be about $\unit[20]{Myr}$.
\end{enumerate}

We believe that the FBs are most likely the hot interior of the EBs and shine in gamma rays presumably by inverse Compton up-scattered photons due to relativistic electrons, which are most naturally accelerated in the forward shock by the well-known first-order Fermi process. There may have also been a contribution from the reverse shock, when it was still present, but in agreement with the low Mach number of the forward shock it should have retreated by now.

In this work we refrained from performing 3D simulations of the RT instability, as these are computationally much more expensive than their 2D counterparts. However, this pragmatic decision does not change the fact that turbulence, to which the RT instability ultimately leads, is an inherently three-dimensional phenomenon. For the non-linear regime, this additional dimension can be taken into account by modifying the value of the potential energy release rate coefficient $\alpha$. But although single modes grow faster in 3D than in 2D, this is not necessarily described by a higher value of $\alpha$, since this is determined by several other factors besides the number of dimensions \citep[including the ``connectivity'' of the interpenetrating structures; see][]{Dim:04}. Nonetheless, all the values of $\alpha$ found so far for the three-dimensional RT instability from both simulations \citep[cf.][]{Sch:20} and experiments \citep[cf.][]{Ban:20} should lie within the broad range considered here. As far as the interstellar shells and supershells are concerned, 2D treatment should be justified as the thickness of the RT unstable layer is small compared to the radius of curvature of the bubbles.

It is important to note that applying the analytical methods outlined in this paper is much faster than performing full-blown hydrodynamical simulations of interstellar bubbles and SBs, which have to provide sufficiently high spatial resolutions, particularly in the usually rather thin shells and supershells, where the RT instability develops. This is complicated by the fact that the widely-used schemes of Godunov-type are notorious for smearing stationary and slowly-moving contact discontinuities \citep{Tor:09}. For instance, the second-order piecewise-parabolic method was shown to require at least 25 grid cells per wavelength to capture the growth rate of the RT instability correctly \citep{Cal:02}.

We close by stressing that the theoretical analysis presented here is quite general and can be applied to problems from solar-system to galaxy-cluster scales.

%%%%%%%%%%%%%%%%%%%%%%%%%%%%%%%%%%%%%%%%%%%%%%%%%%
\section*{Acknowledgements}

We thank Jenny Feige for useful discussions on stellar evolution models and the anonymous referee for comments that helped to improve the manuscript.

%%%%%%%%%%%%%%%%%%%%%%%%%%%%%%%%%%%%%%%%%%%%%%%%%%
\section*{Author Contributions}

M.M.S.~performed and analysed the numerical simulations, designed the required post-processing routines, identified the unstable cases for the linear instability regime, derived analytical perturbation amplitudes for exponential and power-law accelerations, applied the developed methods to astrophysical objects (stellar wind bubbles, young SNRs, and SBs), and wrote the initial draft of the paper. He also produced all figures and curates the research data.

D.B.~kicked off the research, derived the dispersion relation for the linear instability regime, and contributed in extending and revising the manuscript.

%%%%%%%%%%%%%%%%%%%%%%%%%%%%%%%%%%%%%%%%%%%%%%%%%%
\section*{Data Availability}
 
The data underlying this paper will be shared on request to the corresponding author.



\bibliographystyle{mnras}
%\typeout{}
\bibliography{td-rti} % if your bibtex file is called example.bib

%\begin{thebibliography}{99}
%\end{thebibliography}

%%%%%%%%%%%%%%%%%%%%%%%%%%%%%%%%%%%%%%%%%%%%%%%%%%

%%%%%%%%%%%%%%%%% APPENDICES %%%%%%%%%%%%%%%%%%%%%
\comment{
\appendix

\section{Some extra material}

If you want to present additional material which would interrupt the flow of the main paper,
it can be placed in an Appendix which appears after the list of references.
}
%%%%%%%%%%%%%%%%%%%%%%%%%%%%%%%%%%%%%%%%%%%%%%%%%%

% Don't change these lines
\bsp	% typesetting comment
\label{lastpage}
\end{document}